\documentclass{svjour3}
\usepackage{txfonts}
\usepackage{epsfig}
\usepackage{graphicx}
\usepackage{natbib}
\usepackage[margin=1.2in, top=1.2in, bottom=1.2in, a4paper, pdftex, dvips]{geometry}

\newcommand{\Rs}{$R_{\odot}$}
\def\vp{v_{\phi}}
\def\Bp{B_{\phi}}
\def\pa{\partial}

\title{The Sun's interior structure and dynamics, and the solar cycle}

\author{A.-M. Broomhall, P. Chatterjee, R. Howe, A.A. Norton, M.J. Thompson}

\institute{A.-M. Broomhall \at
              Institute of Advanced Studies, University of Warwick, Coventry, CV4 7HS, UK \\
              Centre for Fusion, Space, and Astrophysics, Department of Physics, University of Warwick, Coventry CV4 7AL, UK \\
              Tel.: +44-2476-574211\\
              Fax: +44-2476-150897\\
              \email{a-m.broomhall@warwick.ac.uk}           
            \and
            P. Chatterjee \at
              High Altitude Observatory, National Center for Atmospheric Research, PO Box 3000, Boulder CO 80307, USA \\
              \email{mppiyali@ucar.edu}            
            \and
              R. Howe \at
              School of Physics and Astronomy, University of Birmingham, Edgbaston, Birmingham, B15 2TT, UK \\
              \email{rhowe@noao.edu}               
            \and
              A.A. Norton \at
              HEPL, Solar Physics, Stanford University, CA 94305 USA  \\
              \email{aanorton@stanford.edu}               
            \and
              M.J. Thompson \at
              High Altitude Observatory, National Center for Atmospheric Research, PO Box 3000, Boulder CO 80307, USA \\
              \email{mjt@ucar.edu}               
}

\begin{document}
\maketitle

\begin{abstract}
The Sun's internal structure and dynamics can be studied with helioseismology, which uses the Sun's natural acoustic oscillations to build up a profile of the solar interior. We discuss how solar acoustic oscillations are affected by the Sun's magnetic field. Careful observations of these effects can be inverted to determine the variations in the structure and dynamics of the Sun's interior as the solar cycle progresses. Observed variations in the structure and dynamics can then be used to inform models of the solar dynamo, which are crucial to our understanding of how the Sun's magnetic field is generated and maintained.
\end{abstract}

\section{Introduction to helioseismology}\label{section[intro]}

Helioseismology is the study of the solar interior using observations of waves that propagate within the Sun. The Sun's natural acoustic resonant oscillations are known as solar p modes.  The p stands for pressure as the main restoring force is a pressure differential.  At any one time thousands of acoustic oscillations are traveling throughout the solar interior. Each individual solar p mode is
trapped in a specific region of the solar interior, known as a cavity, and its
frequency is sensitive to properties, such as temperature and
mean molecular weight, of the solar material in the cavity. The sound waves can be considered as damped harmonic oscillators as they are stochastically excited and intrinsically damped by turbulent convection in the outer approximately 30\,per cent by radius of the solar interior. The strongest p-mode oscillations have a periodicity of approximately 5\,minutes. The solar oscillations can be observed in two ways: by line-of-sight Doppler velocity measurements over the visible disk; or by measuring the variations in the continuum intensity of radiation, which are caused by the compression of the radiating gas by the waves.

The horizontal structure of p modes can be modeled as spherical harmonics and so can be described by three main components: Harmonic degree, $l$, which indicates the number of node lines on the surface; azimuthal degree, $m$, which describes the number of planes slicing through the equator and takes values between $-l\le m\le l$; and radial degree, $n$, which gives the number of radial nodes in the solar interior. If the Sun was completely spherically symmetric the frequencies of the oscillations would be degenerate in $m$. However, asphericities in the solar interior, such as rotation, temperature variations and the presence of magnetic fields, lift the degeneracy and measurements of the differences in the frequencies of the $m$ components allow inferences to be made concerning the effects responsible for the asphericities.

The frequency of a mode, $\nu_{n,l,m}$, can be expressed in terms of the $m=0$ central frequency, $\nu_{n,l}$, and a polynomial expansion of splitting (or $a$) coefficients, $a_j(n,l)$:
\begin{equation}\label{equation[splitting]}
    \nu_{n,l,m}=\nu_{n,l}+\sum^{j_{\textrm{\scriptsize{max}}}}_{j=1}a_j(n,l)\mathcal{P}_j^{(l)}(m),
\end{equation}
where $\mathcal{P}_j^{(l)}(m)$ are the Ritzwoller-Lavely formulation of the Clebsch-Gordon expansion \citep{1991ApJ...369..557R} and are given by
\begin{equation}\label{equation[RL]}
    \mathcal{P}_j^{(l)}(m)=\frac{l\sqrt{(2l-j)!(2l+j+1)!}}{(2l)!\sqrt{2l+1}}C^{lm}_{j0lm}.
\end{equation}
Here, $C^{lm}_{j0lm}$ are the Clebsch-Gordon coefficients. The odd-order $a$ coefficients express the difference in mode frequency caused by the modes' interaction with the rotation profile of the solar interior. The even-order coefficients are sensitive to all other departures from spherical symmetry. These could include local variations in the sound speed or cavity size \citep{1996MNRAS.278..437B}, temperature variations \citep{1988ApJ...331L.131K}, second order rotation effects from the differential rotation \citep{1992ApJ...394..670D}, and the presence of magnetic fields \citep{1988IAUS..123..175G}. Isolating the contributions of all these effects to the even-order coefficients is extremely difficult.

As sound waves travel from their excitation point near the surface into the solar interior they are refracted because of increasing pressure, and therefore sound speed (since $c_s^2=\frac{\gamma_1P}{\rho}$, where $\gamma_1$ is the first adiabatic exponent). Assuming the direction of travel is not radial the waves follow a curved trajectory which takes them back to the surface where they are reflected by the sharp drop in density. By the time the waves reach the surface they are approximately traveling in a radial direction and so the depth at which the near-surface
reflection of the waves occurs depends only on the wave frequency.
The p modes are set up by a superposition of such waves, and so the location of the upper
edge of the acoustic cavity of the modes -- the upper turning point --
likewise depends only on frequency,
with the upper turning point of low-frequency modes being deeper in the solar interior than the upper turning point of high-frequency modes. The radius of the lower turning point depends on the angle of trajectory and therefore $l$, with low-$l$ modes turning deeper within the solar interior than high-$l$ modes.

Helioseismology can be split into two categories: global and local. Global helioseismology studies the natural resonant acoustic oscillations of the solar interior that are able to form standing waves in the entire Sun, as opposed to local helioseismology, which studies propagating waves in part of the Sun. We now give a brief introduction to each category.

\subsection{Global helioseismology}\label{section[global intro]}

Global helioseismology can itself be split into two sub-categories: Sun-as-a-star (unresolved) observations; and resolved observations. Sun-as-a-star observations are only sensitive to the lowest harmonic degrees ($0\le l\le 3$, and occasionally $l=4$ and 5). These modes are the truly global modes of the Sun as they travel right to the energy-generating core. However, by making resolved observations of the solar surface spatial filters can be applied that allow measurements of modes with degrees into the 100s or even 1000s.

The Birmingham Solar Oscillations Network \citep[BiSON, ][]{1995A&AS..113..379E, 1996SoPh..168....1C} has been making Sun-as-a-star line-of-sight velocity observations for over 30 years, covering cycles 21 (although with limited coverage), 22, 23, and 24. BiSON is an autonomous network of 6 ground-based observatories strategically positioned around the world so as to allow observations of the Sun to be made 24\,hr a day. This is important as gaps in the time domain contaminate frequency spectra, making it more difficult to accurately obtain the parameters that describe the modes, such as their frequencies. One can also make continuous observations of the Sun from space and this is the approach taken by the SOlar and Heliospheric Observatory (SOHO) spacecraft, which was launched in 1995. Sun-as-a-star observing programs onboard SOHO include the Global Oscillations at Low Frequencies \citep[GOLF; ][]{1995SoPh..162...61G} instrument, which also measures line-of-sight velocity, and the Variability of solar IRrandiance and Gravity Oscillations \citep[VIRGO; ][]{1995SoPh..162..101F}, which measures the changes in intensity of the Sun.

VIRGO is also capable of making resolved observations of the Sun, as was the Michelson Doppler Imager \citep[MDI; ][]{1995SoPh..162..129S}, also onboard SOHO. Global Oscillations Network Group (GONG) is a ground-based network of 6 sites \citep{1996Sci...272.1284H}, which has been making resolved observations of the Sun since 1995. More recently, the Helioseismic and Magnetic Imager \citep[HMI; ][]{2012SoPh..275..229S} and the Atmospheric Imaging Assembly \citep[AIA; ][]{2012SoPh..275...17L} onboard the Solar Dynamics Observatory (SDO) have been used for helioseismic studies \citep{2011JPhCS.271a2061H, 2011JPhCS.271a2058H}. SDO was only launched in 2010 and so these instruments cannot be used to study previous solar cycles. They are, however, likely to play an important role in helioseismic studies of cycle 24, and possibly beyond.

\subsection{Local helioseismology}\label{section[local intro]}

Since the advent of helioseismic observations at high spatial resolution,
a number of different data analysis techniques known collectively as
``local helioseismology'' have been developed. They do not rely on the
global resonant-mode nature of the observed oscillations: rather, the
observations of wave motions are interpreted in terms of their local
properties. Three methods in particular are widely used. Ring analysis (or
ring-diagram analysis) works within a freqeuncy-wavenumber framework for
analysing the oscillations, albeit on localized patches of the solar surface,
and may therefore be considered the local-helioseismic analogue of
global-mode helioseismology. Time-distance helioseismology (or
helioseismic tomography),
and the closely related method of acoustic holography,
work rather in the framework of interpreting
the effect of heterogeneities and flows on propagating waves in terms of
travel-time shifts or equivalently phase shifts. These two basically
distinct approaches will be described briefly below.

In ring analysis \citep{1989ApJ...343L..69H},
measurements of
wave properties are made in localized regions (``tiles'') on the surface of
the Sun. The observable - typically the Doppler velocity - over the pixels
within the region are Fourier transformed in the two spatial horizontal
directions and in time to give wave power as a function of spatial
wavenumbers $k_x$, $k_y$ and frequency. In cuts through the 3-D power spectrum
at fixed frequency, the power is found to lie in rings corresponding to the
different radial orders $n$ of the modes: hence the name ``ring analysis''.
In the absence of flows or any horizontal inhomogeneities, the dispersion
relation of the waves is of the form $\omega = \omega_n(k_h)$,
where $k_h \equiv \sqrt{k_x^2+k_y^2}$ is the magnitude of the horizontal
wavenumber vector. In other words, the frequencies do not depend on the
direction of the horizontal wavenumber, only on its magnitude: thus, the rings
of power are circular and centered on $(k_x,k_y)=(0,0)$. Large-scale flows
and magnetic fields, which can change the dispersion relation depending
on the direction of propagation of the waves, shift and distort the rings.
Further, local changes in the effective isotropic wave speed (such as could
be caused by thermal anisotropies) change the frequency at fixed wavenumber
or equivalently change the radius of the rings at fixed frequency. These
various perturbations to the rings, measured at different frequencies and
for different $n$ values, permit the possibility of performing a
one-dimensional inversion in depth to infer the large-scale flow, wave speed,
etc., below each tile \citep{2002ApJ...570..855H, 2008SoPh..251..439B, 2010ApJ...713L..16G}. By combining inversion results under different tiles,
a three-dimensional subsurface map can be built up. Recently, fully 3-D
inversion methods have been developed and implemented that simultaneously
invert the data from many different tiles to obtain 3-D maps of subsurface
flows directly \citep{2011JPhCS.271a2002F}.

In time-distance helioseismology,
travel times of waves that propagate beneath the surface between pairs of
surface points are
estimated by cross-correlating the observed oscillations between surface
points. The cross-correlation function derived by cross-correlating the
oscillations at two surface points A and B, say, exhibits a number of
``wave packets''. In a ray-theoretic interpretation, these packets correspond
to waves that travel between A and B along different ray paths, either
traveling directly without intermediate surface bounces, or bouncing at the
surface one or more times before arriving at the point of observation. In
practice, it can be necessary to average the cross-correlations between
many similarly separated pairs of points before the wave packets are clearly
visible . Partly for this reason, the observations are typically filtered in
some way before the cross-correlations are made. The most common filters
are a phase-speed filter or a filter that aims to isolate oscillations corresponding
to a particular radial order $n$. Also, usually it is only the first-arrival
wave packet -- corresponding to propagation between A and B with no
intermediate surface bounces -- that is used in the subsequent analysis.
Travel times are estimated by measuring the location of the wave packet, and
a number of different methods are used to do that. Flows, magnetic fields
and inhomogeneities experienced by the waves during their propagation from A
to B shift the location of the wave packet (and this in turn is measured as
a travel-time shift) as well as potentially modifying the width and amplitude
of the wave packet. Also, flows can cause a difference in travel time
depending on whether the waves are traveling from A to B or from B to A.
By measuring travel times between many different pairs of surface points,
3-D subsurface maps of flows and wave-speed inhomogeneities can be made
using inverse techniques \citep{2004ApJ...603..776Z}.

For further discussion of the signatures of flows on the ring analysis and
time-distance helioseismology measurements, see Section~\ref{sec:2.1}. For now, however, we return to concentrate on the global modes.

\section{An introduction to global helioseismology and the solar cycle}\label{section[global intro]}
\begin{figure}
  \centering
  \includegraphics[clip, width=0.6\textwidth, trim=2cm 1cm 4cm 19cm]{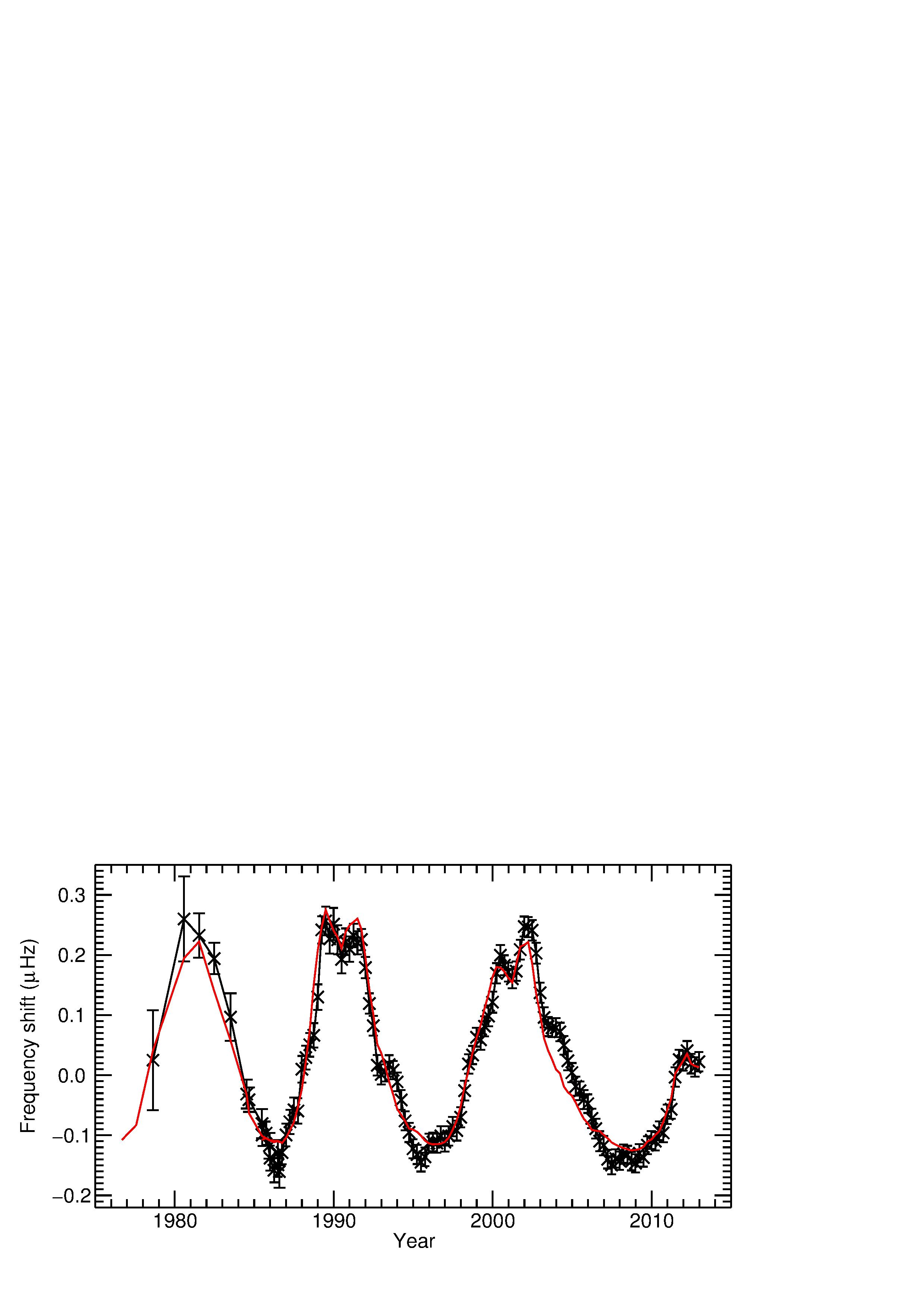}\\
  \caption{Average frequency shifts of ``Sun-as-a-star'' modes with
  frequencies between 2.5 and $3.5\,\rm mHz$. The results were
  obtained from 365\,d BiSON time series that overlapped by 91.25\,d. Also plotted is a
  scaled and shifted version of the 10.7\,cm flux.}
  \label{figure[BiSON shifts]}
\end{figure}

It has been known since the mid 1980s that p-mode frequencies vary
throughout the solar cycle with the frequencies being at their
largest when the solar activity is at its maximum
\citep[e.g.][]{1985Natur.318..449W, 1989A&A...224..253P, 1990Natur.345..322E,
1990Natur.345..779L, 2007ApJ...659.1749C, 2007ApJ...654.1135J}.\footnote{Please note here
that the references listed are not exhaustive, instead we have tried
to include useful ones whose reference lists themselves are
informative.} For a low-$l$ mode at about $3000\,\rm\mu
Hz$ the change in frequency between solar maximum and minimum is
about $0.4\,\rm\mu Hz$. By examining the changes in the observed
p-mode frequencies throughout the solar cycle we can learn about
solar-cycle-related processes that occur beneath the Sun's surface. The 11 year cycle is seen clearly in Fig. \ref{figure[BiSON shifts]}, which shows the mean frequency shifts of the
p modes observed by BiSON and the 10.7cm flux\footnote{http://www.ngdc.noaa.gov/stp/space-weather/solar-data/solar-features/solar-radio/noontime-flux/penticton/} for comparison \citep[also see][]{2009ApJ...700L.162B,
2009A&A...504L...1S, 2010ApJ...718L..19F}.

What causes the observed frequency shifts? The
magnetic fields can affect the modes in two ways: directly and indirectly. The direct effect occurs because the Lorentz force provides an additional restoring force resulting in an increase
of frequency, and the appearance of new modes. Magnetic fields can influence the oscillations indirectly by affecting the physical
properties of the cavities in which the modes are trapped and, as a result, the propagation of
the acoustic waves within them. For example, changes in the thermal structure can affect both the propagation speed and the location of the upper turning point. Indirect effects can both increase or decrease the frequencies of p modes.

To date the relative contributions from the direct and indirect effects remain uncertain. \citet{2005ApJ...625..548D}
suggest that the magnetic fields are too weak in the near-surface layers for
the direct effect to contribute significantly to the observed
frequency shifts and that the indirect effects dominate the perturbations. However, \citeauthor{2005ApJ...625..548D} also suggest that the direct effect may be more important for low-frequency modes deeper within the solar interior where the magnetic field is strong enough to produce a noticeable shift in frequency. These
results are still controversial and appear to disagree with the results of \citet{1986Natur.323..603R}, who found that field strengths of the order of 500\,kG would be required at the base of the convection zone to explain the shift. \citet{2005A&A...439..713F} investigated the effect of a magnetic field at the base of the convection zone and a more shallow magnetic field, at a depth of 50\,Mm, but found their direct influence on the frequencies of p modes was consistently smaller than the change in frequency that is observed. Changes in both the chromospheric magnetic field strength and temperature can explain the observed downturn in the magnitude of the frequency shifts at high frequencies \citep{1993ApJ...414..898J}. However, as we now describe, it is not only the mode frequencies that vary throughout the solar cycle.

\subsection{Solar cycle variations in mode powers and lifetimes}

Solar p modes are excited and damped by turbulent convection beneath
the solar surface. The process of excitation and damping varies
throughout the solar cycle and these variations are observed as
changes in the heights and widths of the mode profiles in a
frequency-power spectrum. For typical low-$l$ modes, as the surface
activity increases the mode frequencies and widths increase but the
mode heights decrease. An increase in the linewidths implies that
the modes experience more damping at times of high activity and so
the lifetimes decrease.

Numerous authors have observed that lifetimes decrease with solar
activity and those observations have been made using data from
different instrumental regimes and for a range of $l$
\citep[e.g.][]{1990LNP...367..135J, 1990LNP...367..129P, 2000MNRAS.313...32C, 2000ApJ...543..472K, 2001ESASP.464...71A, 2001ESASP.464..123T,
2002SoPh..209..247J, 2003ApJ...588.1204H, 2004ApJ...604..969J, 2006ApJ...650..451S, 2009SoPh..258....1B, 2010A&A...516A..30S}.
The size of the variation appears to be dependent on $l$ with
lower-$l$ modes showing a larger percentage change than high-$l$
modes \citep[e.g.][and references therein]{2009SoPh..258....1B}.
\citet{2009SoPh..258....1B} looked at the lifetimes of modes with
$300\le l\le600$ and found that in active regions the lifetimes
decrease as activity increases by about 13\% between minimum and
maximum. In quiet regions the lifetimes still decrease with solar
activity but to a lesser extent (the lifetimes at solar maximum are
about 8\% of the lifetimes at solar minimum). This implies that the
change to the damping is not just associated with the active regions
that appear on the surface. In general p-mode linewidths increase with frequency, except for a
plateau region at around $2800\,\rm\mu Hz$. \citet{2000ApJ...543..472K} found
that the linewidths of modes in the plateau region were most
sensitive to the level of solar activity. 

Mode powers have been observed to decrease with increasing solar activity \citep[e.g.][]{1990LNP...367..129P, 1992A&A...255..363A, 1993MNRAS.265..888E}. The decrease in mode power is of the order of
20\% and has been now been observed by many authors
\citep[e.g.][]{2000MNRAS.313...32C, 2000ApJ...543..472K, 2001A&A...379..622J, 2001ESASP.464...71A, 2001ESASP.464..123T, 2002SoPh..209..247J, 2004ApJ...604..969J, 2009ASPC..416..281S}. The total energy of the mode, $E$, can be determined by multiplying
the total power of the mode by the mode mass \citep[as defined by][]{1991sia..book..401C}. The mode energy shows a similar
decrease to the mode power, with a maximum to minimum variation of
approximately 12\% \citep{2002SoPh..209..247J}.

The energy supply rate is determined by multiplying the mode energy
by the mode width \citep[e.g.][]{2000MNRAS.313...32C}. Numerous studies have
shown that while the mode energy decreases with activity the energy
supply rate shows no solar cycle variation
\citep[e.g.][]{2000MNRAS.313...32C, 2000ApJ...543..472K, 2001ESASP.464...71A, 2001ESASP.464..123T, 2002SoPh..209..247J, 2003ApJ...588.1204H, 2004ApJ...604..969J, 2006ApJ...650..451S}. Therefore, the
observed variations in mode energy and mode power probably arise
from an increase in damping only. As the energy of the modes
decreases between solar minimum and solar maximum this implies that
some energy has gone missing. 

One potential source of damping is magnetic activity on the solar
surface. Active regions are known to suppress the power of p modes, and effect lifetimes, and energy supply rates \citep[e.g.][]{1981SoPh...69..233W, 1982ApJ...253..386L, 1992ApJ...394L..65B, 2001ApJ...563..410R, 2002ApJ...572..663K, 2004ApJ...608..562H}. However, the associated mechanisms are not yet fully
understood. One explanation is that strong-field magnetic regions,
such as sunspots, are effective absorbers of p-mode power
\citep[e.g.][]{1999ApJ...515..832H, 2002A&A...387.1092J}. \citet{2000ApJ...543..472K} suggested that the
energy could be in flux tubes, whose numbers increase with solar
activity, and the energy could excite oscillations in magnetic
elements. Another suggestion is that the
efficiency of mode excitation is reduced in magnetic areas
\citep[e.g.][]{1988ApJ...326..462G, 1995ApJ...451..372C, 1996ApJ...464..476J}. Other
possible explanations as to why p-mode excitation is suppressed in
sunspots include a different height of spectral line
formation due to the Wilson depression or a modification of p-mode
eigenfunctions by the magnetic field \citep[][and references therein]{2009SoPh..258....1B}.

Another possibility that could alter p-mode damping is variations in
the convective properties near the solar surface, which are most
likely to arise from the influence of magnetic structures \citep{2001MNRAS.327..483H}. \citeauthor{2001MNRAS.327..483H}
theorized that changes of parameters in the convection zone would
affect linewidth shapes mainly in the plateau region of a frequency
spectrum, as is observed. This, therefore, implies that during times of
high-magnetic activity the convection zone is affected sufficiently
to produce a measurable change in p-mode linewidths. Several authors
have observed that the horizontal size of solar granules decreases
from solar minimum to solar maximum \citep[e.g.][]{1984ssdp.conf..265M,
1988AdSpR...8..159M, 1999A&A...344..965B, 2007A&A...475..717M}. \citet{1988AdSpR...8..159M} found that the horizontal granule size decreased by approximately 5\% from solar minimum to solar maximum. According
to \citet{2001MNRAS.327..483H} this should result in an increase in damping
rates of about 20\%, which is in reasonable agreement with the
observed change in damping rates in BiSON data of approximately
$24\pm3$\%.

We now return to look in more detail at the changes in frequencies of the p modes with solar cycle.

\section{Dependence of solar cycle frequency shifts on $l$ and frequency}

\begin{figure}
\centering
  \includegraphics[width=0.45\textwidth, clip, trim=2cm 1cm 3cm 17cm]{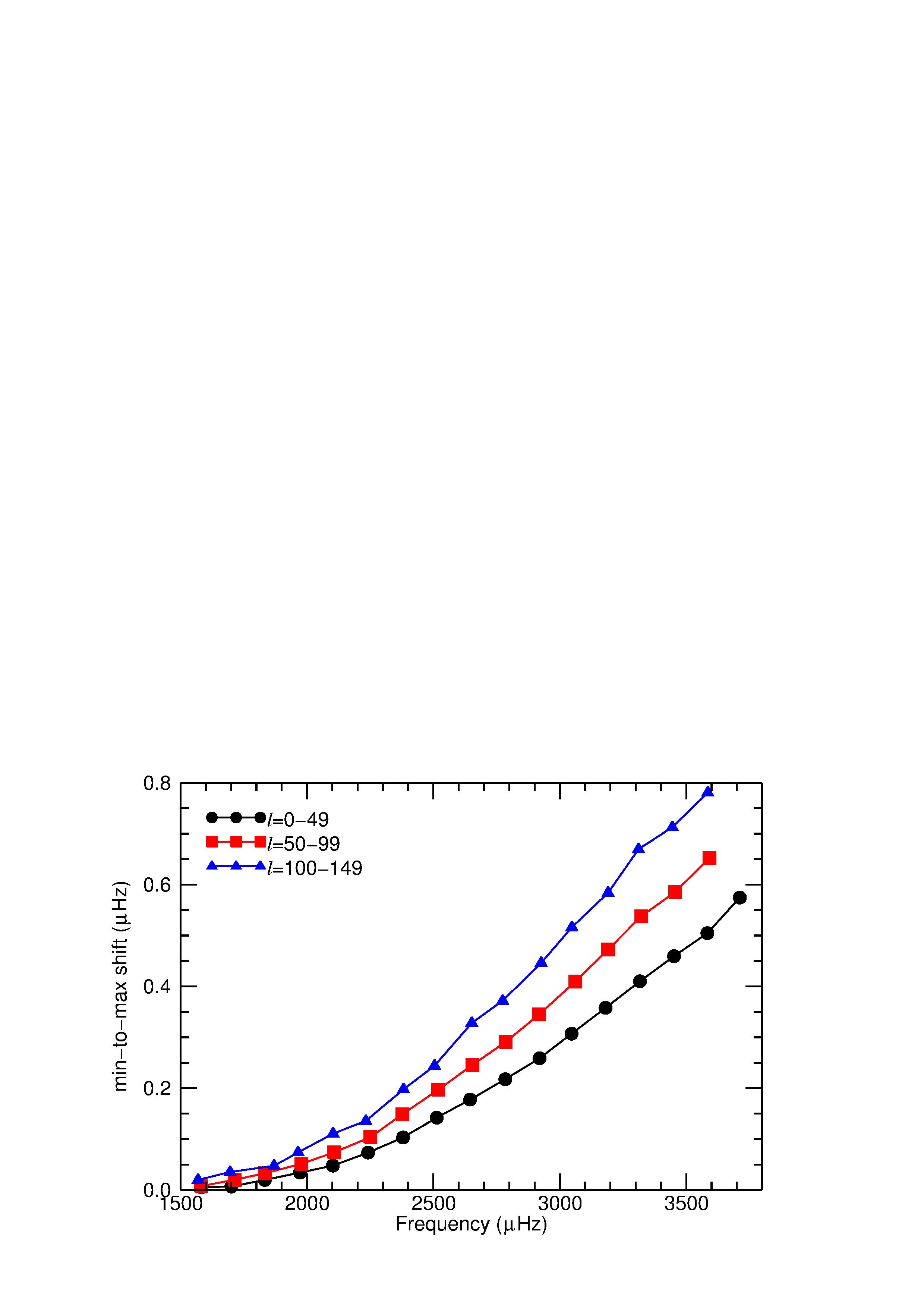}
  \includegraphics[width=0.45\textwidth, clip, trim=2cm 1cm 3cm 17cm]{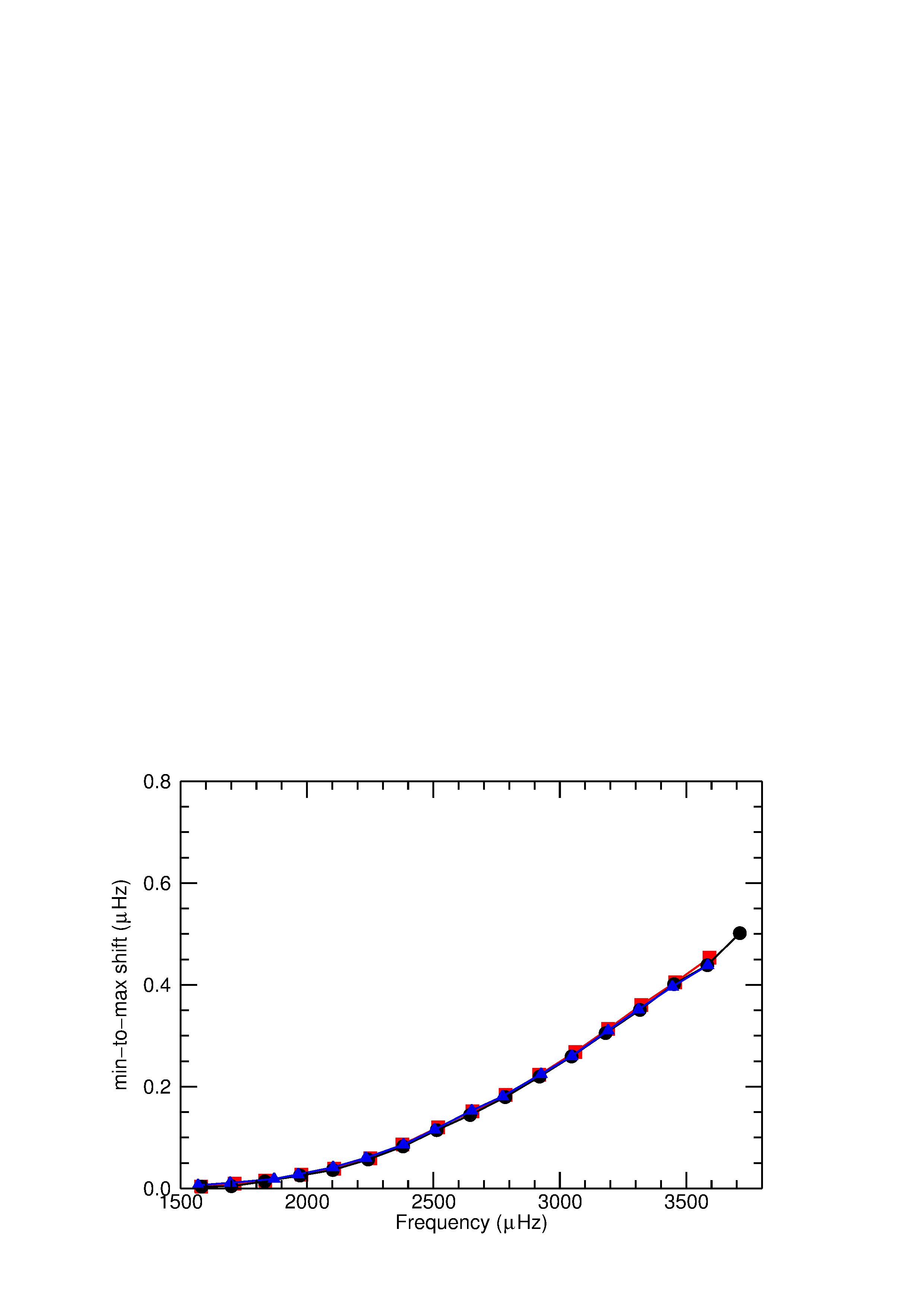}\\
  \caption{Left-hand panel: Change in frequency between solar minimum and solar maximum, $\delta\nu$, as a function of mode frequency. The results have been obtained by averaging over different ranges in $l$ (see legend), and by smoothing over $135\,\mu Hz$. Right-hand panel: Frequency shifts that have been scaled by the inertia ratio, $Q_{n,l}$. The errorbars on these figures are not visible because they are smaller than the symbol size. Results obtained using GONG frequencies. }\label{figure[shift vs freq]}
\end{figure}

Solar cycle frequency shifts, $\delta\nu_{n,l}$, have well-known
dependencies on both angular degree, $l$, and frequency,
$\nu_{n,l}$, as demonstrated in Fig. \ref{figure[shift vs freq]} \citep[also see e.g.][]{1990Natur.345..779L, 1994ApJ...434..801E, 1998MNRAS.300.1077C,
1999ApJ...524.1084H, 2001MNRAS.324..910C, 2001A&A...379..622J}. We now look in more detail at these dependencies and
discuss what they tell us about the origin of the perturbation. We note here that the frequency shifts plotted in Fig. \ref{figure[shift vs freq]} were determined for the central frequency of the mode i.e. the $m=0$ frequency. In section \ref{section[surface comp]} we discuss the latitudinal dependence of the shifts and there we consider modes with different $m$.

\subsection{Dependence of solar cycle frequency shifts on mode inertia}
The main $l$-dependence in $\delta\nu_{n,l}$ is associated with
mode inertia. The normalized mode inertia is defined by \citet{1991sia..book..401C} as
\begin{equation}\label{equation[mode inertia]}
   I_{n,l}=M_\odot^{-1}\int_{v}|\xi|^2\rho\textrm{d}V=
   4\pi M_{\odot}^{-1}\int_0^{R_s}|\xi|^2\rho r^2\textrm{d}r=
   \frac{M_{n,l}}{M_{\odot}},
\end{equation}
where $\xi$ is the displacement associated with a mode, suitably
normalized at the photosphere, $V$ is the volume of the Sun, and
$M_{\odot}$ is the mass of the Sun. $M_{n,l}$ is the `mass'
associated with a mode. The physical interpretation of the mode
inertia is some measure of the interior mass affected by any
given mode. At fixed frequency a decrease in $l$ results in an
increase in $M_{n,l}$ and therefore in $I_{n,l}$. In other words as $l$ decreases a greater volume of the
interior is associated with the motions generated by the mode.
Therefore the high-$l$ modes are more sensitive to a perturbation and so vary more throughout the
solar cycle than low-$l$ modes.

The inertia ratio, $Q_{nl}$, is defined by \citet{1991sia..book..401C} as:
 \begin{equation}
 Q_{nl} = I_{nl} / \bar{I}(\nu_{nl}),
 \label{eqution[inertia ratio]}
 \end{equation}
$\bar{I}(\nu_{nl})$ is the inertia an $l=0$ modes would have at a frequency $\nu_{nl}$. Multiplying the frequency shifts by $Q_{nl}$ removes the $l$ dependence of the frequency shifts at fixed frequency, as can be seen in Fig. \ref{figure[shift vs freq]}. This makes the dependence a function of frequency alone. Collapsing the $l$ dependence in this manner allows the frequency shifts of a wide range of $l$ to be combined, thereby reducing any uncertainties associated with the shifts and allowing tighter constraints to be placed on the frequency dependence.

\subsection{Dependence of solar cycle frequency shifts on mode frequency}
\label{section[freq dependence of shifts]}

The frequency dependence of the frequency shifts, which can be seen in the right-hand panel of Fig. \ref{figure[shift vs freq]}, is a telltale indicator that the
observed 11-year signal must be the result of changes in acoustic
properties in the few hundred kilometres just beneath the visible
surface of the Sun, a region to which the higher-frequency modes are
much more sensitive than their lower-frequency counterparts
because of differences in the upper boundaries of the cavities in
which the modes are trapped \citep{1990Natur.345..779L, 1991sia..book..401C}. We now
go into more detail.

As mentioned earlier, for a given $l$ the upper turning point of low-frequency modes is
deeper than the upper turning point of high-frequency modes. At a
fixed frequency lower-$l$ modes penetrate more deeply into the solar
interior than higher-$l$ modes. Therefore, higher-frequency modes,
and to a lesser extent higher-$l$ modes, are more sensitive to
surface perturbations.

\citet{1990Natur.345..779L} discuss the origin of the perturbation. If the
perturbations were to extend over a significant fraction of the
solar interior, asymptotic theory implies that the fractional mode frequency
shift would depend mainly on $\nu_{n,l}/l$, which is not what we observe. This implies that the relevant structural changes
occur mainly in a thin layer. \citet{1988ESASP.286..321T} found that the
effect of perturbing a thin layer in the propagating regions of
modes, such as the layer where the second-ionization of helium occurs, is an oscillatory frequency
dependence in $\delta\nu_{n,l}$. Such a frequency dependence is also not observed, implying
that the dominant frequency dependence is not the direct result of,
for example, changes in the magnetic field at the base of the
convection zone. If the perturbation was confined to the centre of
the Sun the size of the frequency shift would increase with
decreasing $l$ as low-$l$ modes penetrate deeper into the solar
interior than high-$l$ modes. In fact, $\delta\nu_{n,l}$ increases
with increasing $l$.

We can think of the frequency dependence as a power law where $\delta\nu_{n,l}\propto\nu_{n,l}^\alpha$. A perturbation from a layer strictly confined to the photosphere (but extending over less than one pressure scale height) is
expected to result in an $\alpha=3$
relationship \citep{1980tsp..book.....C, 1990LNP...367..283G, 1990Natur.345..779L,
1991ApJ...370..752G}. If instead the perturbation
extends beneath the surface, the frequency dependence will be weaker
and $\alpha$ will get smaller \citep{1990LNP...367..283G}. \citet{2001MNRAS.324..910C} determined that $\alpha<3$, which  implies that the significant contribution comes from a perturbation close to the surface, in the
sub-photospheric layers. \citet{2008AdSpR..41..861R} observed that $\alpha$ is smaller for the
lower-frequency modes, which suggests that the perturbation extends
to greater depths the lower in frequency one goes. This is
consistent with the results of \citet{2005ApJ...625..548D} (see Section \ref{section[global intro]}).

\begin{figure}
  \centering
  \includegraphics[width=0.45\textwidth, clip, trim=4cm 10cm 4cm 9cm]{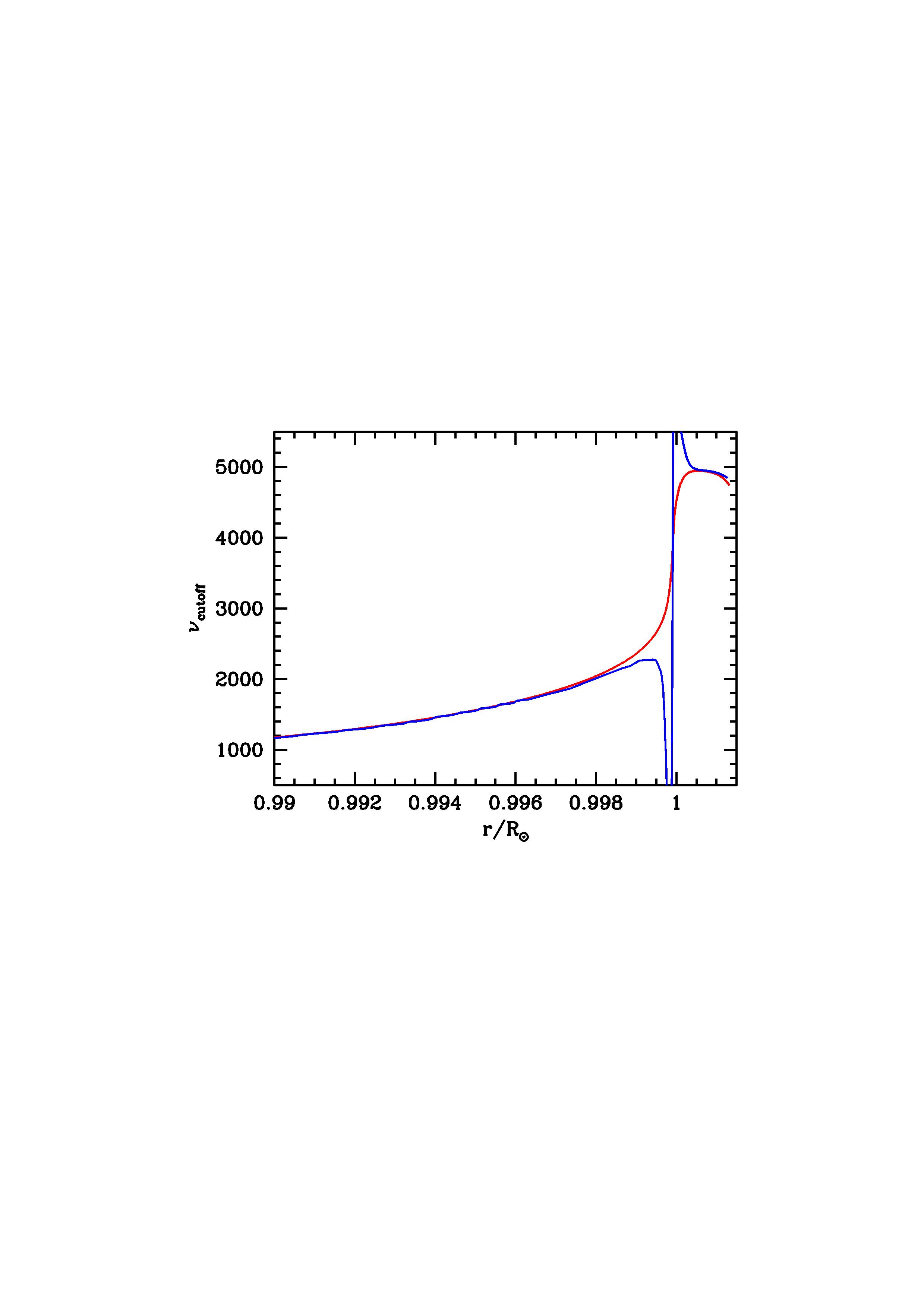}\\
  \caption{Acoustic cut-off frequency as a function of solar radius. The upper turning point of the modes can be thought of in terms of the acoustic cut-off frequency because the upper turning point corresponds to the radius at which the acoustic cut-off frequency equals the mode frequency. The blue curve shows the cut-off frequencies determined from a standard solar model, while the red curve shows the acoustic cut-off frequency in an isothermal limit. Adapted from \citet{2012ApJ...758...43B}.}
  \label{figure[upper turning points]}
\end{figure}

In summary, the oscillations are
responding to changes in the strength of the solar magnetic activity
near the Sun's surface. As modes below approximately $1800\,\rm\mu
Hz$ experience almost no solar cycle frequency shift it is
reasonable to conclude that the origin of the perturbation is
concentrated in a region above the upper turning points of these
modes. Fig. \ref{figure[upper turning points]} shows that the
upper turning point of a mode with a frequency of $\sim1800\,\rm\mu
Hz$ is about $0.996R_\odot$ (approximately 3\,Mm) below the surface. Note
that the upper turning point predicted by a model is strongly
dependent on the properties of the model at the top of the
convection zone.

Above $3700\,\rm\mu Hz$ the relationship between frequency shift and
activity changes, as the magnitude of the solar cycle frequency
shift decreases and oscillates \citep[e.g.][]{1990Natur.345..779L,
1991ApJ...370..752G, 1996ApJ...456..399J}. Furthermore, above $\sim4100\,\rm\mu Hz$
it appears that modes experience a decrease in frequency as activity
increases \citep{1994SoPh..150..389R, 1998MNRAS.300.1077C}.

As we have seen the frequency dependence of the frequency shifts implies that the shifts can be associated with a near-surface perturbation.  Therefore, we now move on to directly compare the change in mode frequency with the magnetic field that is observed at the solar surface (and beyond into the solar atmosphere).

\subsection{Comparison between frequency shifts and the surface magnetic field}\label{section[surface comp]}

Numerous comparisons have been made between the p-mode frequency shifts and various proxies of the Sun's surface magnetic field \citep[e.g.][]{1998A&A...329.1119J, 2001MNRAS.327.1029A, 2001SoPh..200....3T, 2002ApJ...580.1172H, 2004MNRAS.352.1102C, 2007ApJ...659.1749C, 2009ApJ...695.1567J, 2012A&A...545A..73J}, including the 10.7\,cm and the International Sunspot Number\footnote{http://www.ngdc.noaa.gov/stp/space-weather/solar-data/solar-indices/sunspot-numbers/international/} (ISN). \citet{2007ApJ...659.1749C} compared six different proxies with the low-$l$ mode frequency shifts and demonstrated that better correlations are observed with proxies that measure both the strong and the weak components of the Sun's magnetic field, such as the Mg II H and K core-to-wing data, the 10.7 cm radio flux, and the He I equivalent width data (as opposed the the ISN and the Kitt Peak Magnetic Index, which mainly sample the strong magnetic flux). The weak-component of the solar magnetic flux is distributed over a wider range of latitudes than the strong component and so a possible explanation of these results is in terms of the latitudinal distribution of the modes used in this study (i.e. low-$l$ modes).

\begin{figure}
  \centering
  \includegraphics[width=0.45\textwidth, clip,  trim=2cm 1cm 3cm 17cm]{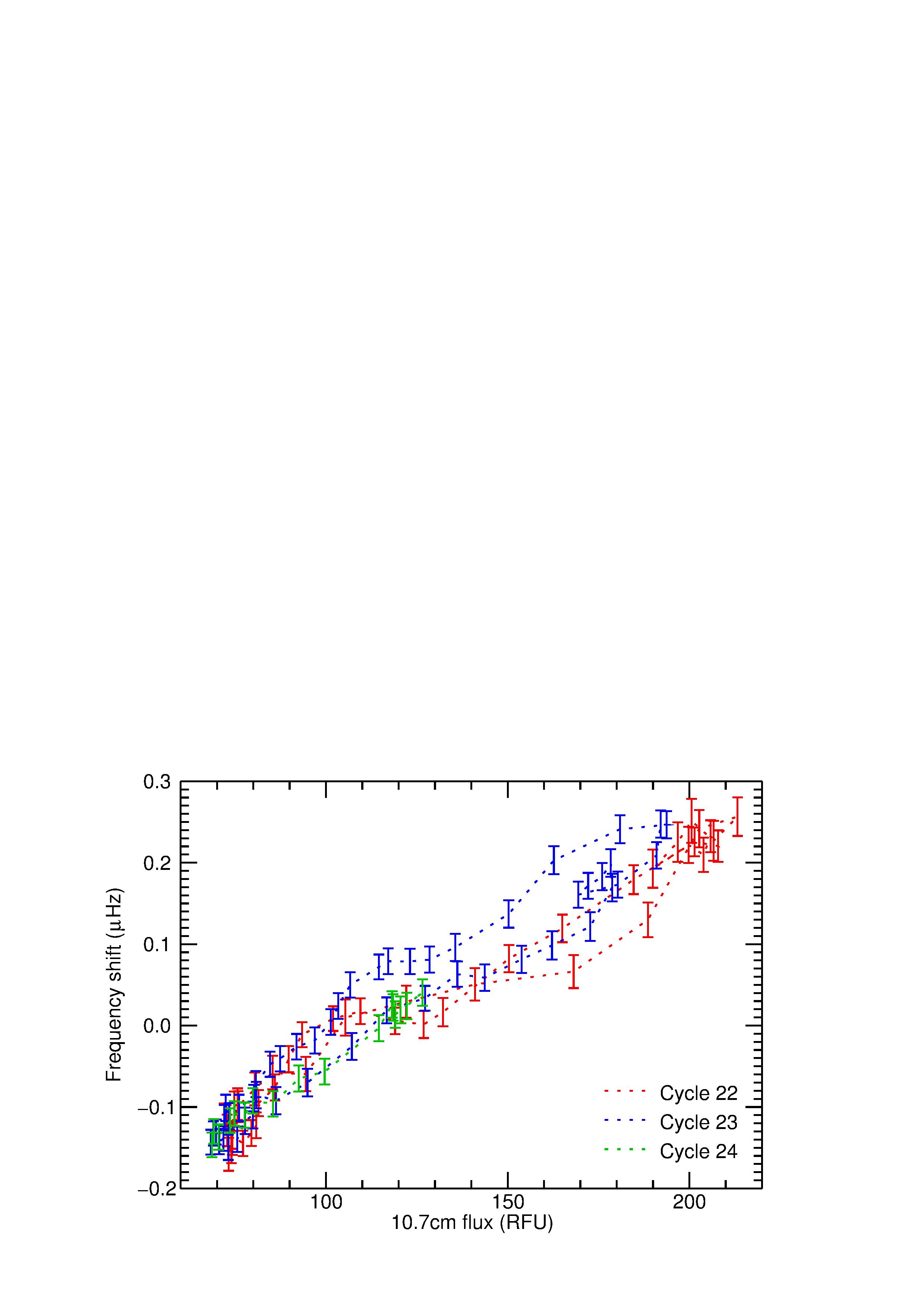}\hspace{1cm}\includegraphics[width=0.45\textwidth, clip,  trim=2cm 1cm 3cm 17cm]{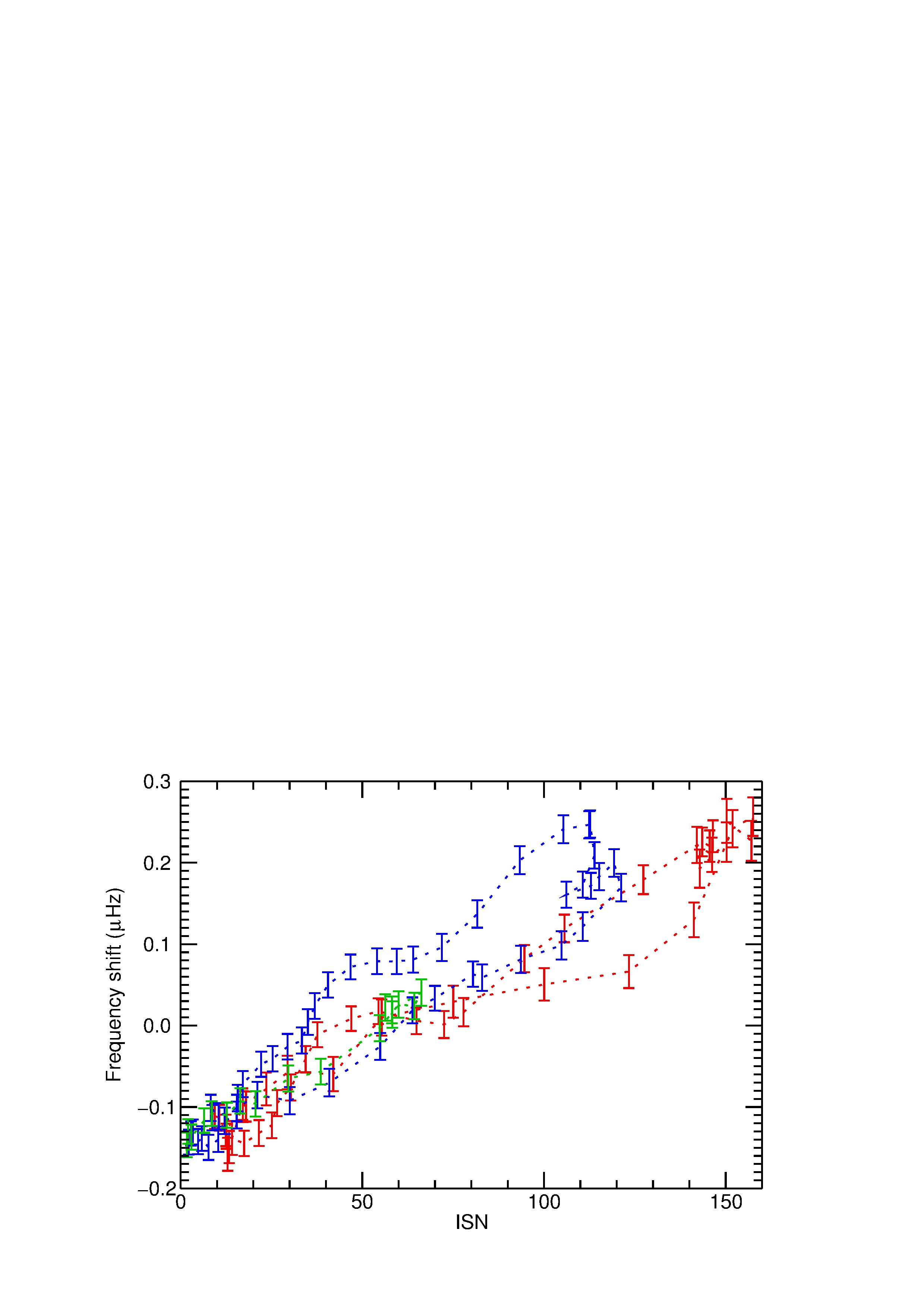}\\
  \caption{Comparison of frequency shifts with surface proxies of the Sun's magnetic field. The left-hand panel shows a comparison with the 10.7\,cm flux, while the right-hand panel shows a comparison with the ISN. The different activity cycles are indicated by different colours (see legend).}\label{figure[proxy vs shift]}
\end{figure}

Fig. \ref{figure[proxy vs shift]} shows a comparison between two proxies of the Sun's magnetic field and low-$l$ frequency shifts. Although the agreement is approximately linear the well-known hysteresis is clearly visible, particularly in the case of the sunspot number. This can be explained in terms of the variation in the latitudinal distribution of the surface magnetic field throughout the solar cycle \citep{2000MNRAS.313..411M}: The different modes have different latitudinal dependencies, as described by the appropriate spherical harmonics (see Section \ref{section[intro]}), and so the relative influence of the surface magnetic flux on the modes changes throughout the solar cycle as the strong surface magnetic field migrates towards the equator.

\begin{figure}
\centering
  \includegraphics[width=0.6\textwidth, clip, trim=2cm 13cm 1cm 4cm]{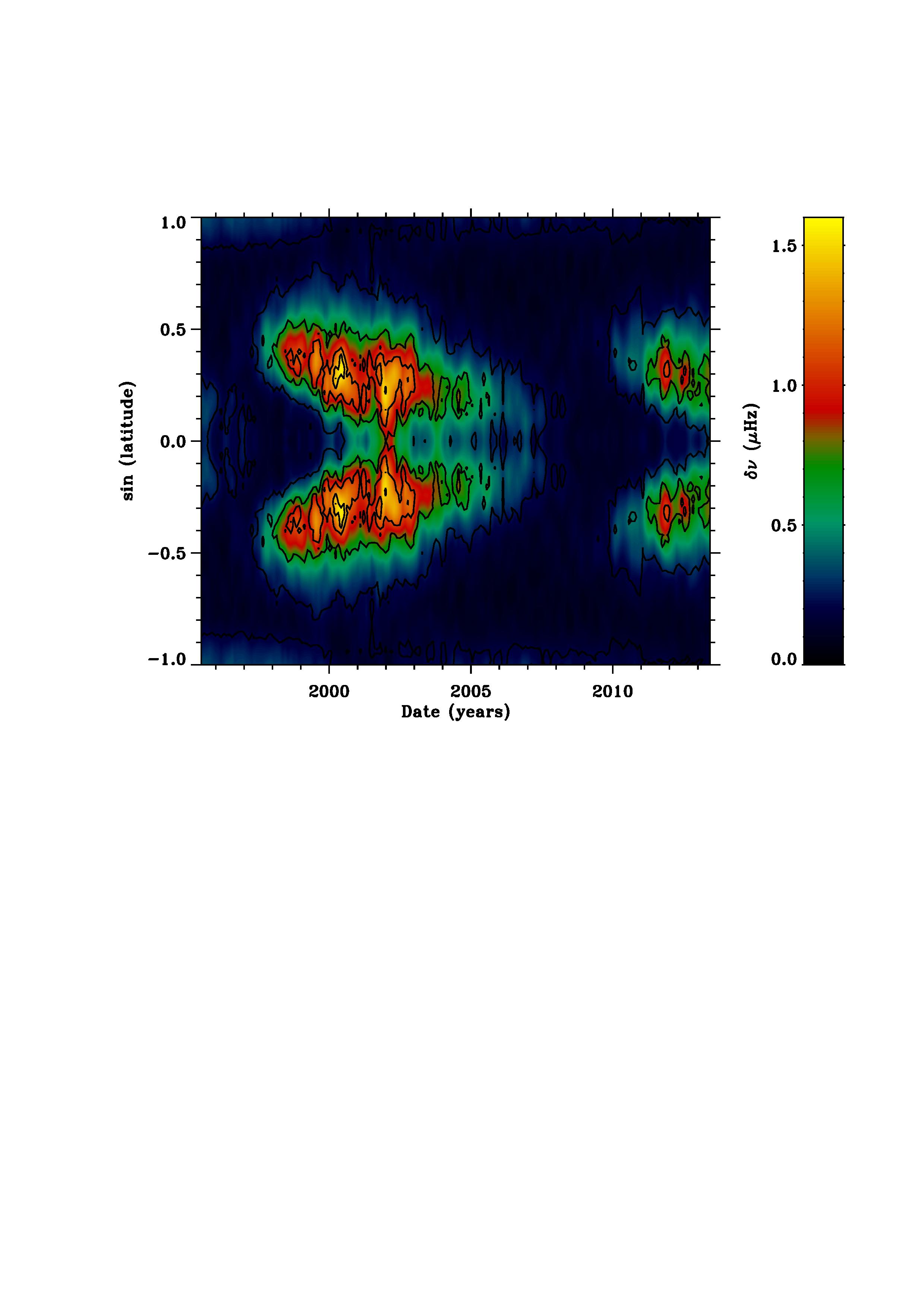}\\
  \caption{Frequency shift inversions using GONG data and modes with $40\le l <80$, and $9\le n \le 11$. The size of the shift is indicated by the colour and the black contours show the surface magnetic field at 5\,G intervals. Updated from \citet{2002ApJ...580.1172H}.}\label{figure[shift_butterfly]}
\end{figure}

As mentioned in Section \ref{section[intro]} the even splitting coefficients can provide information about departures from spherical symmetry, such as those produced by the presence of a magnetic field. Although the correlation between the even $a$ coefficients and global measures of the Sun's magnetic field is good, it is not linear. However, the even splitting coefficients are linearly correlated with the corresponding Legendre polynomial decomposition of the surface \citep[e.g.][]{1999ApJ...524.1084H, 2002ApJ...580.1172H, 2003MNRAS.343..343C, 2004A&A...424..713C,
2004MNRAS.352.1102C, 2004ApJ...610L..65J} indicating that the size of the observed frequency shift experienced by a particular mode is dependent on the latitudinal distribution of the surface magnetic flux. Furthermore, \citet{2002ApJ...580.1172H} showed that it is possible to use latitudinal inversion techniques to localize the frequency shifts in latitude and, in fact, reconstruct the evolution of the surface magnetic field. An up-to-date version of these inversions can be seen in Fig. \ref{figure[shift_butterfly]}, which clearly resembles the familiar butterfly diagrams usually associated with the surface magnetic field.

\section{Evidence for structural changes due to magnetic fields deeper in the interior}

All that has been discussed above concerns the near-surface magnetic field. However, one of the main advantages of helioseismology is that it allows the deeper interior to be studied. Therefore we now move on to discuss attempts to detect evidence of the solar magnetic field deeper in the solar interior. However, this is understandably hard. The plasma-$\beta$ $(=P_\textrm{\scriptsize{gas}}/P_\textrm{\scriptsize{mag}})$ in the deep interior is significantly greater than unity. Furthermore, the sound speed in the interior increases substantially with depth, meaning the modes spend significantly longer in the near-surface regions than in the deep interior. All this means that the influence of a deep-seated magnetic field on the properties of the oscillations will be limited. However, the rewards for finding evidence of deep-seated magnetic fields are great. For example, many believe the solar dynamo is generated in the tachocline at the base of the convection zone \citep[see][for a recent review]{2010LRSP....7....3C}.

Numerous authors have looked but were unable to find helioseismic evidence for magnetic fields deeper in the solar interior \citep[e.g.][]{1988IAUS..123..155G, 2000SoPh..192..449B, 2000SoPh..192..481B, 2001MNRAS.327.1029A, 2001MNRAS.324..498B, 2002ESASP.508....7B, 2003ESASP.517..231B, 2002ApJ...580..574E, 2003ApJ...591..432B}. They therefore resorted to putting limits on parameters at the base of the convection zone, such as a maximum change in field strength between solar minimum and maximum of 300\,kG \citep{1997MNRAS.288..572B, 2000A&A...360..335A} or a change in sound speed of $\delta c/c=3\times10^{-5}$ \citep{2002ApJ...580..574E}.

\citet{2005ApJ...624..420C} and \citet{2005ApJ...633.1187S} considered the frequency shift, scaled by mode mass, as a function of the horizontal phase speed, which is given by
\begin{equation}\label{equation[horiz phase speed]}
    w=\frac{\nu}{2\pi[l(l+1)]^{1/2}},
\end{equation}
and so can be related to the lower turning point of the mode. The authors use both MDI and GONG data and use $190\le w \le 1570$. They observed that the scaled frequency shift was approximately constant with horizontal phase speed around solar minimum. However, as the surface magnetic field increased, the scaled frequency shift decreased, but only above a critical horizontal phase speed value, which corresponds to a depth near the base of the convection zone. They interpret this as indicating that the wave speed near the base of the convection zone changes with activity and they find that this behaviour is consistent with a magnetic perturbation at the base of the convection zone. Further, they find that $\delta c/c=1-3\times10^{-5}$, which implies a change in magnetic field of between 170-290\,kG. These results are consistent with the upper limits set by earlier authors \citep{1997MNRAS.288..572B, 2000A&A...360..335A, 2002ApJ...580..574E}.

\citet{2008ApJ...686.1349B} and \citet{2009ASPC..416..477B} used a principal component analysis (PCA), which reduces the dimensionality of the data and consequently the noise, to find a small but statistically significant change in the frequencies of modes whose lower turning points are at or near the base of the convection zone. This change is tightly correlated with surface activity. If the change in frequency can be interpreted as a change in sound speed due to the presence of a magnetic field \citet{2009ASPC..416..477B} find that their results imply a change in field strength between the maximum of cycle 23 and the preceding minimum of 390\,kG, just above the limits set previously.

Moving slightly closer to the surface, \citet{2004ApJ...617L.155B} showed evidence for solar structure changes around the zone associated with the second ionization of helium (approximately $0.98R_\odot$). A spherically symmetric localized feature or discontinuity in the internal structure of the Sun causes a characteristic oscillatory component in mode frequencies \citep[e.g.][]{1988IAUS..123..151V, 1990LNP...367..283G, 1994MNRAS.267..209B, 1994MNRAS.268..880R}. \citet{2004ApJ...617L.155B} considered changes in the oscillatory signal caused by the zone in which the second ionization of helium occurs, using intermediate degree modes whose lower turning points were below the depth of the second ionization zone of helium but above the base of the convection zone, which is a discontinuity of its own right. \citeauthor{2004ApJ...617L.155B} found changes in the amplitude of the oscillatory signal and these variations scaled linearly with the surface magnetic field. They explained this in terms of changes in the equation of state, since magnetic fields contribute to both energy and pressure. These results were verified by \citet{2006ApJ...640L..95V} using Sun-as-a-star data, and so using only low-$l$ modes. However, we note that \citet{2011MNRAS.414.1158C} found no evidence for a variation in the amplitude of the oscillatory signal.

Further evidence for changes in the solar interior came from \citet{2012ApJ...745..184R}, who used the frequency differences of intermediate- and high-degree modes observed between the solar cycle 23 maximum and the preceding minimum to infer changes in the relative sound speed squared. \citeauthor{2012ApJ...745..184R} found that the sound speed is larger at solar maximum than at solar minimum at radii greater than $0.8R_\odot$ and that the difference in sound speed increases with radial position from about $0.8R_\odot$ until about $0.985R_\odot$. Below $0.8R_\odot$ the uncertainties are too large and there is too much spurious variation introduced by the inversion process \citep{1996MNRAS.281.1385H} to definitively state whether the sound speed is greater at solar minimum or maximum. However, at its peak $(\sim0.985R_\odot)$ the relative difference in sound speed squared is of the order of $10^{-4}$, which is statistically significant. Above $0.985R_\odot$ the relative difference in the sound speed decreases with increasing radial position before passing through zero at $\sim0.997R_\odot$ and then becoming negative.
These results can be compared to those obtained using local helioseismology techniques to observe sound speed variations beneath a sunspot. For example, \citet{2008SoPh..251..439B} also observed the change in sound speed beneath a sunspot goes from positive to negative with increasing radial position. Furthermore the locations at which the change in sign were observed to occur were in good agreement. \citeauthor{2012ApJ...745..184R} therefore raises the question over how much of the change in sound speed observed in the global modes is due to local active regions, such as sunspots.

Although determining solar cycle variations in the internal structure of the Sun and uncovering evidence for a deep-seated magnetic field have proved to be very difficult far more success has been attained in measuring changes in the flow fields of the solar interior throughout the solar cycle.

\section{Seismology of flow fields in the convection zone}
\label{sec:2}
\subsection{Observational signatures of flows}
\label{sec:2.1}
Large-scale flows advect the acoustic-gravity waves that propagate in the
interior of the Sun. This gives rise to a number of measurable effects in
helioseismology that can then be used in turn to make inferences about the
properties of the flows. Probably the best-known such effect is that of
rotational splitting of the frequencies of global modes. The first-order
effect on the frequencies is that the frequencies of modes of like $n$ and $l$
but different $m$ are shifted
by an amount that is given by $m$ times a mode-weighted average of the
internal solar rotation rate within the acoustic cavity of the mode. The effect
is primarily due to advection: there is also a Coriolis contribution to the
first-order frequency splitting, but for the observed p~modes the Coriolis
contribution is very small. In the particular case of a star that is rigidly
rotating with uniform rotation rate $\Omega$, the frequency shift experienced
by a mode would be $m(1-C_{nl})\Omega$, where the so-called Ledoux factor
$C_{nl}$ arises from the Coriolis contribution. The frequency
shift due to a more general rotation profile can be written as
$m{\bar \Omega}_{nlm}$
where ${\bar\Omega}_{nlm}$ denotes the mode-weighted average of the
rotation rate. This is the principal effect of large-scale flows on the
frequencies of the global modes. The mode dependence of the averages of the
rotation rate is very useful: it enables inferences to be made about the
spatial variation of the rotation rate, using inversion techniques.

As discussed in Section~\ref{section[local intro]} above, in the
local helioseismic
ring analysis approach,
in the absence of flows or any horizontal inhomogeneities, the dispersion
relation of the waves is $\omega = \omega_n(k_h)$,
where $k_h \equiv \sqrt{k_x^2+k_y^2}$ is the magnitude of the horizontal
wavenumber vector. Thus the frequencies do not depend on the
direction of the horizontal wavenumber, only on its magnitude, and so the rings
of power are circular and centered on $(k_x,k_y)=(0,0)$. In the presence of
a flow, the rings are shifted. Suppose that there is a uniform flow of
speed $U$ in the $x$-direction. Then a Doppler shift of the frequencies
changes the dispersion relation to become
$\omega = \omega_n(k_h) + UK_x$. Provided the flow is weak (in
the sense that $U \ll d\omega_n/dk_h$), the rings are still circular but their
center is shifted in the $-k_x$ direction by an amount proportional to
$U / (d\omega_n/dk_h)$. Thus the direction and magnitude of the ring shift
can be used to infer the direction and magnitude of the flow. Analogously
to the case of global-mode frequency shifts,
for a non-uniform flow (e.g. one that varies with depth) the
shifts will be a mode-dependent average and this allows inferences to be
made about the spatial variation of subsurface flows.

In time-distance helioseismology, in the absence of flows
the travel times in both
directions between surface points A and B should be the same.
Flows advect the waves
and can cause the travel time in one direction (with the flow) to be shorter
than the travel time in the opposite direction. In its simplest terms,
the effect can be understood by thinking about travel times along ray paths:
in one direction the waves travel at a speed $c+{\mathbf U}.{\mathbf s}$,
where $c$ is the sound speed, $\mathbf U$ is the flow speed and $\mathbf s$
is a unit vector along the ray in the direction of travel; whereas
waves travelling in the opposite direction travel at a different speed in
the presence of the flow because they have the sign of $\mathbf s$
reversed. The perturbations to the travel times due to the flow are
measured in time-distance helioseismology and, by cross-correlating
many different pairs of points, the magnitude and direction of subsurface
flows can be inferred.

\subsection{Rotation}
\label{sec:2.2}
\begin{figure}[!ht]
\centering
\includegraphics*[width=8.0cm, clip, trim=3cm 1cm 4cm 18cm]{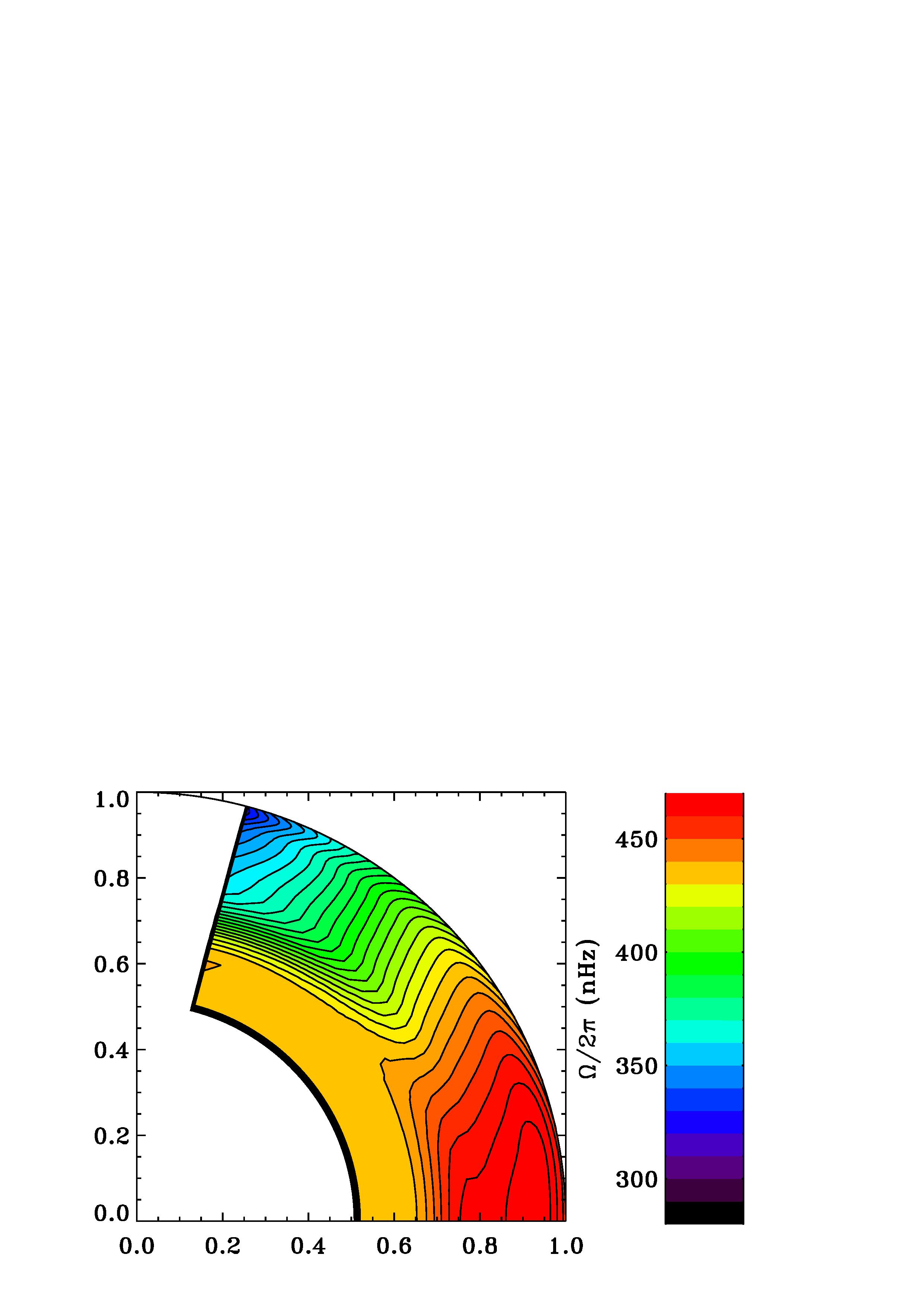}
\caption{Solar internal rotation as inferred from HMI observations using a Regularized Least Squares inversion. Contours of isorotation are shown. \label{HMIrot}}
\end{figure}

The rotation rate in much of the solar interior has been inferred from
global-mode frequency splittings. A typical result for the mean rotation
profile, in this case using a Regularized Least Squares (RLS) inversion
technique applied to data obtained with the HMI instrument on board
SDO, is shown in Fig.~\ref{HMIrot}. The
latitudinal variation of the rotation rate that has long been observed at the
surface of the Sun, with the rotation rate decreasing with increasing latitude,
is seen to persist through the convection zone (the outer 30 per cent of
the Sun). Near the base of the convection zone, there is a transition to
what appears to be an
essentially uniform rotation rate beneath, so that at the interface there is
a region of strong rotational shear which has become known as the
tachocline. There is also a region of rotational shear much closer to the
Sun's surface, in about the outer five per cent by radius, so the maximum
in the rotation rate occurs a few per cent beneath the surface.
As remarked by \citet{2003ESASP.517..283G}, in much of the convection zone the
contours of
isorotation make an angle with the rotation axis of about 25$^\circ$,
whereas there is a
slight tendency for them to align parallel to the rotation axis
(``rotation on cylinders'') in the near-equatorial region \citep{2005ApJ...634.1405H}.

\subsection{Observations of torsional oscillations}
\label{sec:2.3}
\begin{figure}[!ht]
\centering
\includegraphics*[width=9.0cm, clip, trim=1cm 0cm 3cm 21cm]{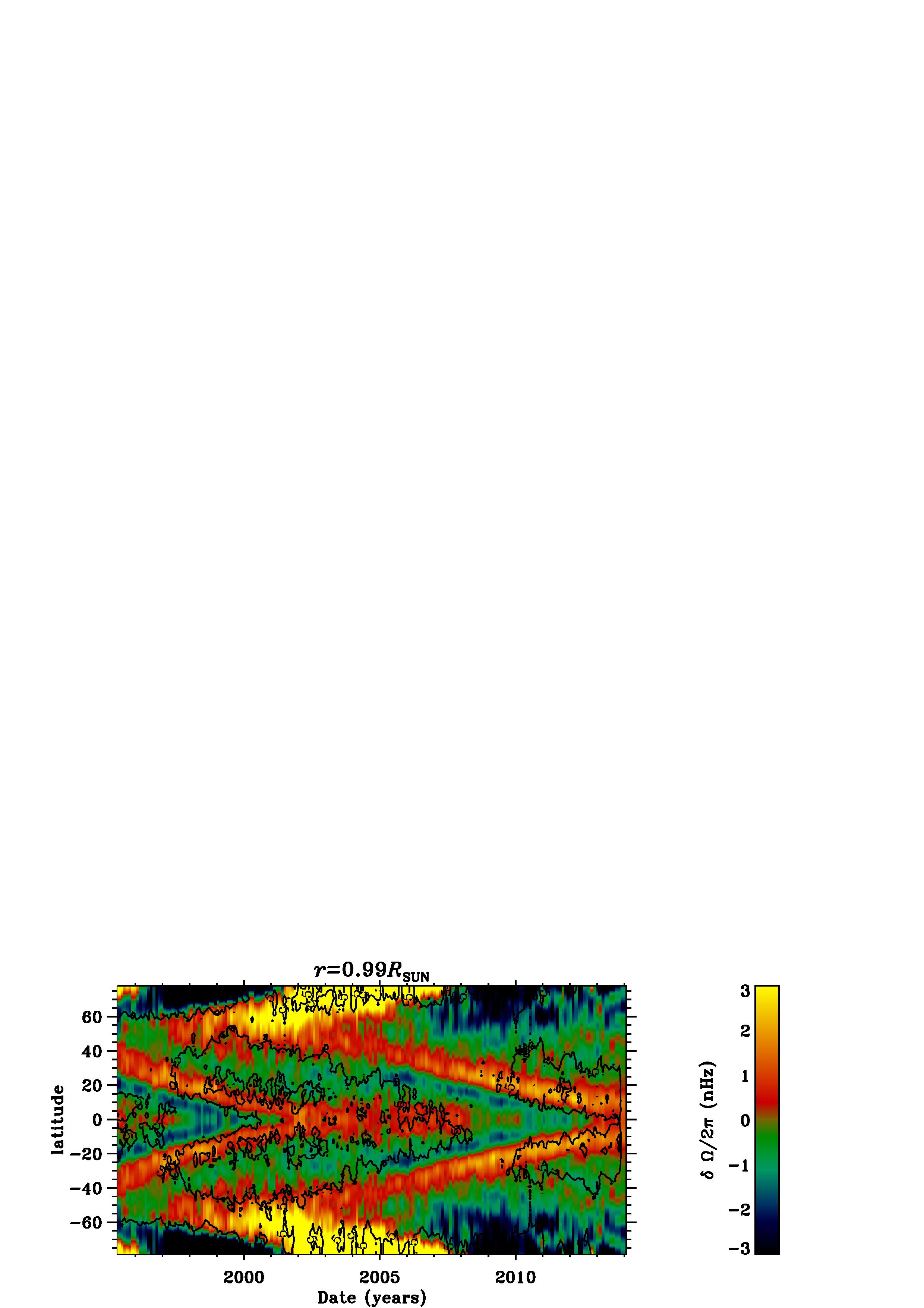}
\caption{Zonal flow residuals after a temporal mean at each radius/latitude has been subtracted. The residuals here are shown as functions of time and
latitude, for a radial location of $0.99R_\odot$. The RLS inversion method was
employed, and the results shown are a merge of inversions of
HMI, MDI and GONG data. The black contours are of the photospheric magnetic
field strength. \label{tors-lat}}
\end{figure}

There are temporal variations around the mean rotation profile. The most
firmly established of these are the so-called torsional oscillations. At low
latitudes, these manifest as weak but coherent bands of faster and slower
rotation that start at mid-latitudes as, or even slightly before, sunspots appear at those latitudes
during the solar cycle, and migrate equatorwards with the activity bands
over a period of a few years. Helioseismology shows that these bands
extend in depth at least a third of a way down into the convection zone \citep{2000ApJ...541..442A,2000ApJ...533L.163H,2002Sci...296..101V,2006ApJ...649.1155H}.
At high latitudes, helioseismology has revealed that there is a
poleward-migrating branch of the torsional oscillation 
\citep{2001ApJ...559L..67A, 2002Sci...296..101V}, which may extend over 
the whole depth of the convection zone. The signal is clearest in the
near-surface layers: Fig.~\ref{tors-lat} shows the torsional oscillations
at a depth of one per cent of the solar radius beneath the surface. As these results are based on global helioseismology they do not reflect any differences
between the flows in northern and southern hemispheres.

The torsional oscillation was first observed in surface Doppler observations from Mount Wilson \citep{1980ApJ...239L..33H}. These observations continued until very recently, and have been compared with the helioseismic observations by
\citet{2006SoPh..235....1H}; when the Doppler observations are symmetrized across the equator the agreement makes it clear that the two techniques are detecting the same phenomenon. The better resolution of the helioseismic measurements at high latitudes makes the poleward-propagating nature of the high-latitude branch more obvious.

\begin{figure}[!ht]
\centering
\includegraphics*[angle=90,width=10.0cm, clip]{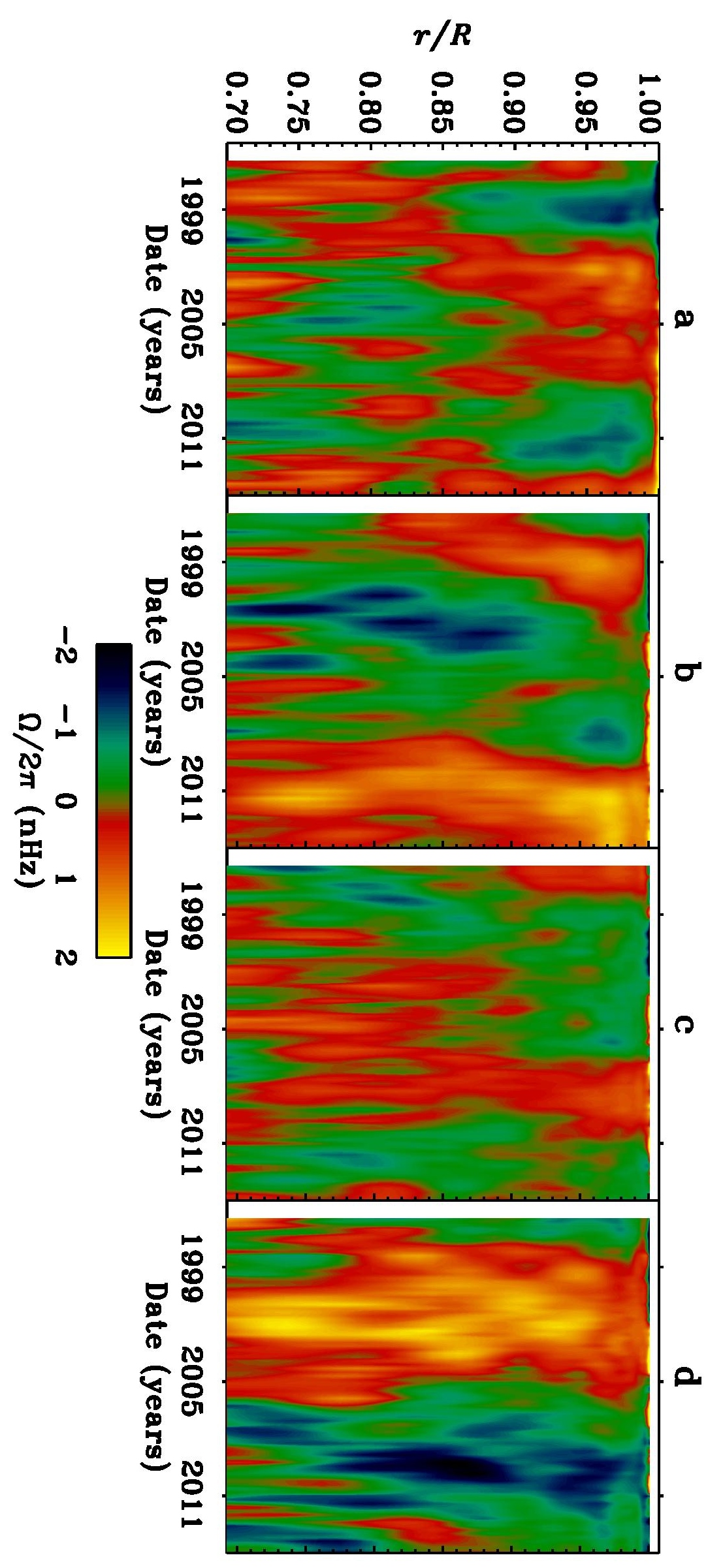}
\caption{Zonal flow residuals from Optimally Localized Averaging inversions of MDI and HMI data. A temporal mean at each radius/latitude has been subtracted. The residuals here are shown as functions of time and
radial location, for latitudes (from left to right) $0^\circ$,
$15^\circ$, $30^\circ$ and $45^\circ$.\label{tors-depth}}
\end{figure}

A complementary view on torsional oscillations is obtained by looking
at the same results but as a function of time and depth at fixed latitude. Four such slices, at different latitudes, are shown in Fig.~\ref{tors-depth}.
At the equator and at $15^\circ$ latitude, there is some hint in
the first half of the time period that the torsional oscillation propagates
upwards from the middle of the convection zone. At the higher
latitudes, there seems to be no propagation in depth. In the second half of
the period there is much weaker evidence for upward propagation, although there
is still some hint of it at $15^\circ$ latitude.

\subsection{Flows around active regions}
\label{sec:2.4}

\begin{figure}[!ht]
\centering
\includegraphics*[width=8.0cm]{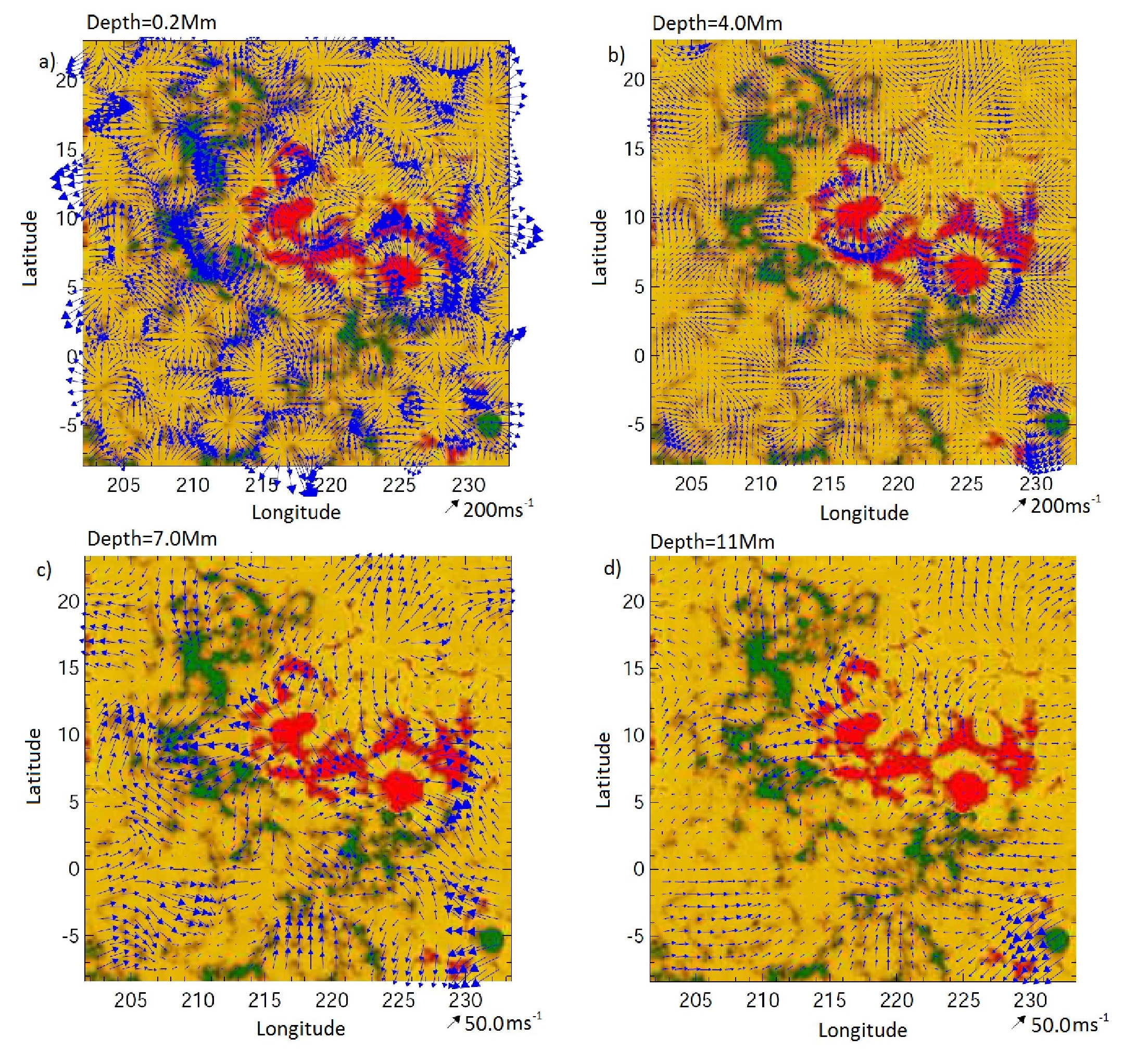}
\caption{Flows around an active region in January 2002, obtained by 3-D
inversion of ring-analysis data using HMI observations. Flows are shown at
depths (a) 0.2~Mm, (b) 4~Mm, (c) 7~Mm and (d) 11~Mm.
Velocity vectors are shown in blue; red and green
show the regions of strong positive and negative polarity photospheric magnetic field, respectively.
 \citep[Adapted from ][]{2011JPhCS.271a2002F}}
 \label{Featherstone1}
\end{figure}

Convection motions manifest themselves at the Sun's surface and in the subsurface regions on a range of scales. Moreover, strong magnetic fields in sunspots
and active regions modify those convective motions, and the
maintenance and decay of sunspots likely involves an interplay between
the magnetic field and the flows. These flows have been studied using various local helioseismic techniques: with time-distance helioseismology \citep{2001IAUS..203..189G}, ring analysis \citep{2009ApJ...698.1749H} and helioseismic holography \citep{2011JPhCS.271a2007B}.
Fig.~\ref{Featherstone1} shows flows around and beneath an active region,
obtained from ring analysis data derived from HMI observations. The inversion
method employed is a 3-D RLS inversion developed by \citet{2011JPhCS.271a2002F}.
At the shallowest depths (0.2 Mm), the flows are
dominated by the supergranular convection, and that scale is clearly visible.
The flow speeds decrease with increasing depth, and the horizontal
scale of the convective motions increases.
Featherstone et al. also find that strong outflows are typical around
sunspots at a depth of 5-6 Mm. Comparable studies with time-distance
helioseismology \citep{2009SSRv..144..249G} agree roughly with the ring-analysis studies in terms of the
magnitude of surface flows near sunspots. They are also in
agreement that there are
strong outflows beneath the surface, but time-distance helioseismology finds
that the flow speeds are higher, by about a factor of two. Resolution
differences may account for some of the apparent discrepancy.

\subsection{Meridional circulation: a shallow return flow?}
\label{sec:2.5}

Meridional circulation is the large-scale flow in meridional planes -- i.e.,
flows that are perpendicular to the rotational flow. Meridional circulation
in the solar convection zone is an important ingredient in some models of the
solar dynamo: in the near-surface
regions it transports the remnant flux from old active regions polewards,
where the flux is presumed to be subducted and carried down to the bottom of the convection zone, where again a suitably directed meridional circulation
may aid the equatorward migration of toroidal magnetic field.

The meridional circulation near the surface of the Sun is only of order
10 m/s, much smaller than the rotation speed and the speeds of convective
motions of granules, etc. Hence, although there were pre-helioseismology
surface measurements of poleward meridional flow at the surface, these
measurements were difficult. In an early application of time-distance
helioseismology,
\cite{1997Natur.390...52G}
demonstrated that the near
sub-surface meridional circulation in both hemispheres is polewards.

Following earlier pioneering attempts by \cite{1995ApJ...455..746P},
Haber and colleagues have applied the ring analysis method extensively to
obtain robust results regarding the meridional circulation in the outer
few percent of the solar interior from the equator to nearly 60$^\circ$
latitude, e.g.
\cite{2002ApJ...570..855H}.
The results show that the meridional circulation is generally poleward in
this region. With a lesser degree of certainty, Haber and collaborators
find that there are episodes when a submerged counter-flow develops
at mid-latitudes.

As well as the possible development of a counter-cell, there are other
temporal variations of the meridional circulation in the superficial
subsurface layers that appear to be associated spatially and temporally
with the torsional oscillations
\citep{2004ApJ...603..776Z,2010ApJ...713L..16G}

\begin{figure}
\includegraphics[width=1.00\textwidth]{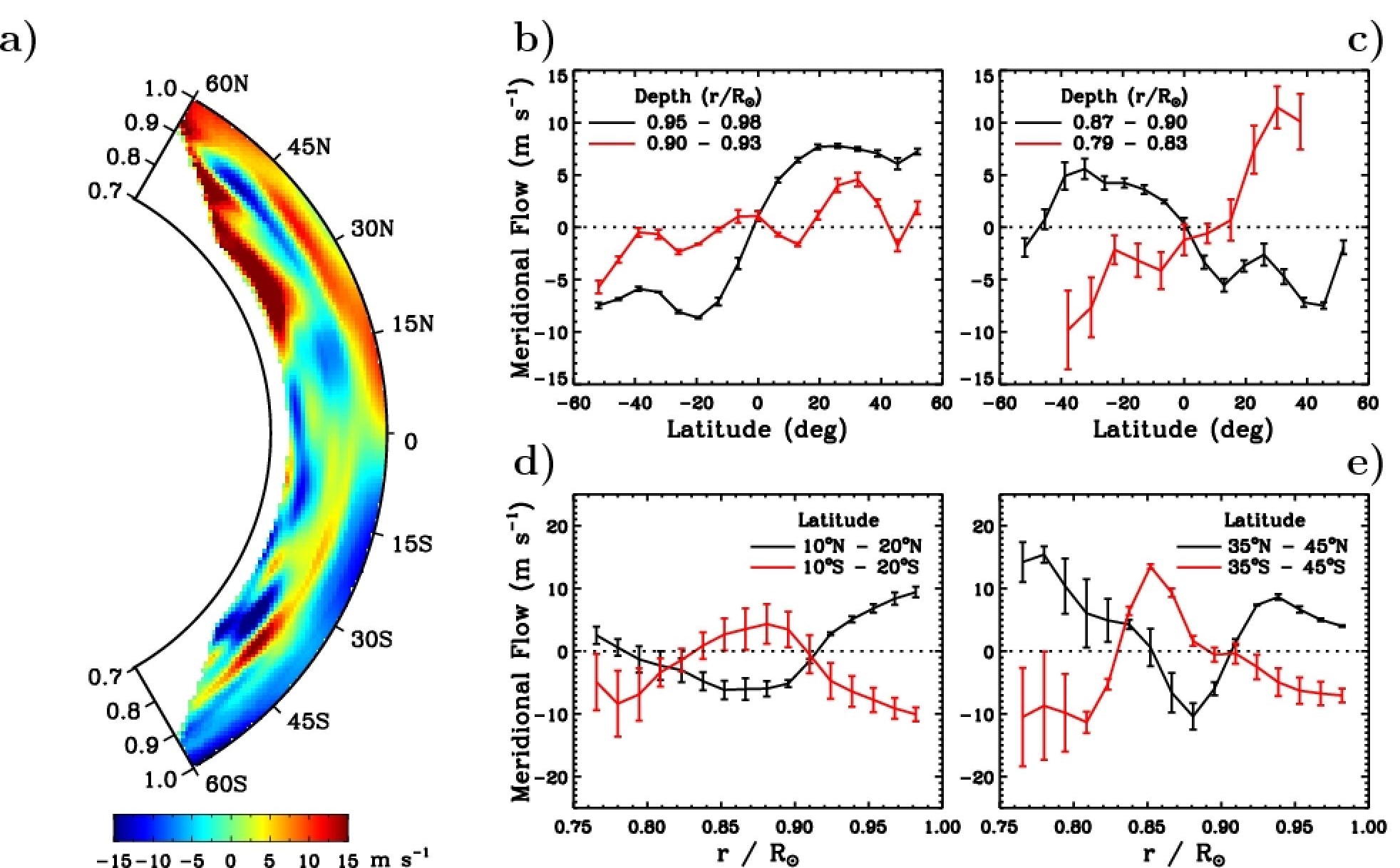}
\caption{The meridional flow profile is shown as obtained by inverting the measured acoustic travel times from two years of HMI data. Panel (a) shows a cross-section view of the profile with the positive velocity directed northward. Panels (b) and (c) show the velocity as functions of latitude averaged over several depths. Panels (d) and (e) show the velocity as functions of depth averaged over different latitudinal bands. From \citet{2013ApJ...774L..29Z}.}
\label{fig:Zhao2013} 
\end{figure}

Mass conservation demands that, corresponding to the poleward near-surface
meridional flow, there must be an equatorward flow at some depth that closes
the circulation. The location of this return flow is a key question in
determining how and whether the flux-transport solar dynamo model operates.
Ever since the work by
\cite{1997Natur.390...52G}, there have been attempts that have hinted at
a relatively shallow return flow. \citet{2007AN....328.1009M}
detected a transition from poleward to equatorward flow at a depth of about
35 Mm, though with substantial error bars on the flow speeds. \citet{2012ApJ...760...84H}, using a cross-correlation method to track
supergranules, found that the poleward meridional flow extends to about
50 Mm beneath the solar surface and reported a positive detection of
equatorward flow at a depth of about 70 Mm.
More recently still, arguably the most convincing helioseismic detection
to date of a shallow return flow is the work by
\cite{2013ApJ...774L..29Z}. These authors find that the poleward flow
extends to about 0.91$R_\odot$ (about 63 Mm depth) with an
equatorward flow between 0.82$R_\odot$ and 0.91$R_\odot$, with perhaps a
poleward flow again below that, see panel a) in Fig.~\ref{fig:Zhao2013}.
However, even with the latest work, a note
of caution is warranted. The result depends on the application of a
correction for a systematic center-to-limb bias in the travel-time
measurements: the correction seems justified from a data-analysis
viewpoint, but as yet the origin of the systematic is not understood.

\subsection{Hemispheric Asymmetry: Flows and Magnetic Field Distribution}
\label{sec:4.1}

The solar cycle appears to be strongly coupled across the equator as evident in the symmetry of the butterfly diagram.  However, a snapshot of the Sun at any given time shows notable differences between the North (N) and South (S).  Usually, the hemispheric asymmetry is most notable in the distribution of the surface magnetic field but it is also observed in flows recovered from local helioseismology.  So the question arises - by what mechanism are the N and S hemispheres coupled?

Passive diffusion across the geometric equator is often considered the main mechanism of coupling \citep{2005SoPh..229..345C}.   Results from dynamo simulations cause us to question magnetic diffusion (including turbulent diffusion) as the main coupling mechanism.  Not only do the diffusion values incorporated into numerical models vary widely, the implementation as a function of depth and the effects of diffusion within the models are not well understood.  Therefore, interest has been increasing regarding other, more active, hemispheric coupling mechanisms.  N-S flows within latitudinally elongated convective cells (aka ``banana cells") allow a mixing of electromagnetic flux from one hemisphere into the other (Passos and Charbonneau, 2014, submitted to A\&A) and can contribute to hemispheric coupling.

Local helioseismology techniques, such as ring diagram \citep{1989ApJ...343L..69H} and time-distance \citep{1993Natur.362..430D} analysis, are able to determine non-symmetric latitudinal structure in the solar interior.   Results from local heliosesmology highlight the differences in the hemispheric flows.  These analyses have been used to measure distinct hemispheric differences in the meridional flows and zonal flows at a given time and depth in the interior \citep[see][and others]{2011JPhCS.271a2077K}.  These measured asymmetries provide further quantitative constraints on the dynamo simulations in that the simulations must reproduce hemispheric asymmetries only within the range observed.

Specifically, the extent of hemispheric coupling as determined by surface magnetism is as follows.  The N and S polar fields reverse their dominant polarity at distinctly different times, up to 14 months apart in some solar cycles \citep{2010SoPh..261..193N}. The time of peak sunspot production of one hemisphere is usually lagged by the other, meaning the hemispheres experience solar maxima at different times.  Recently, the first half of the 2014 calendar year shows the Southern hemisphere producing significantly more large active regions while the Northern hemisphere peaked in sunspot production earlier in 2013.  The amplitudes of each hemispheric solar cycle as measured in sunspot number or area are within 20\% of each other and the timing (phase lags) between hemispheres are not more than 25\% of the total cycle length \citep{2010SoPh..261..193N}.   For more recent, in-depth analysis, \citet{2010AN....331..765Z} and \citet{2013ApJ...765..146M} both analyze hemispheric asymmetry and interestingly enough find that a certain hemisphere appears to lead any given cycle and this lead persists for roughly 40 years on average.  That is the surface magnetism characteristic of the hemispheric asymmetry.

\begin{figure}
\includegraphics[width=0.5\textwidth]{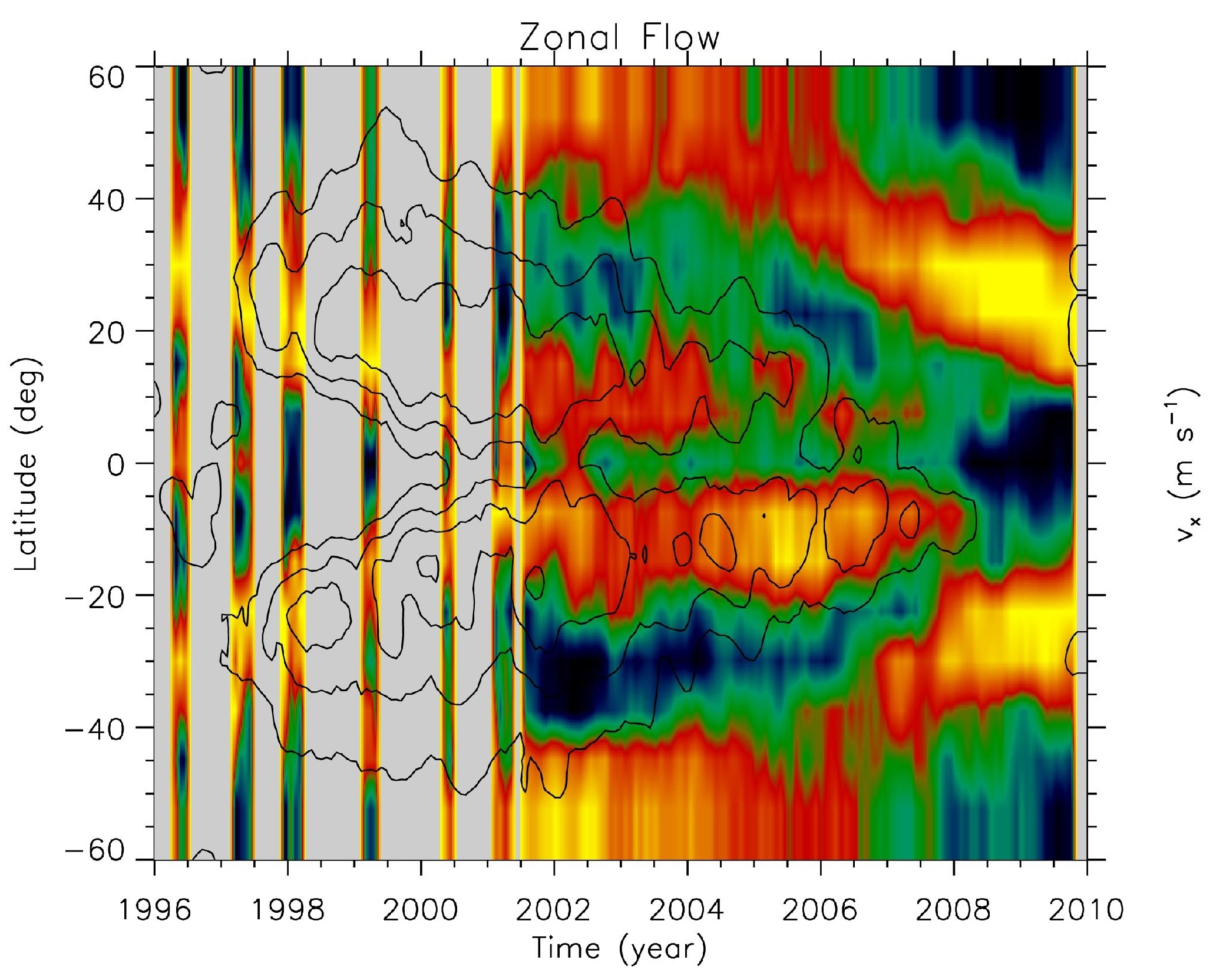}\includegraphics[width=0.5\textwidth]{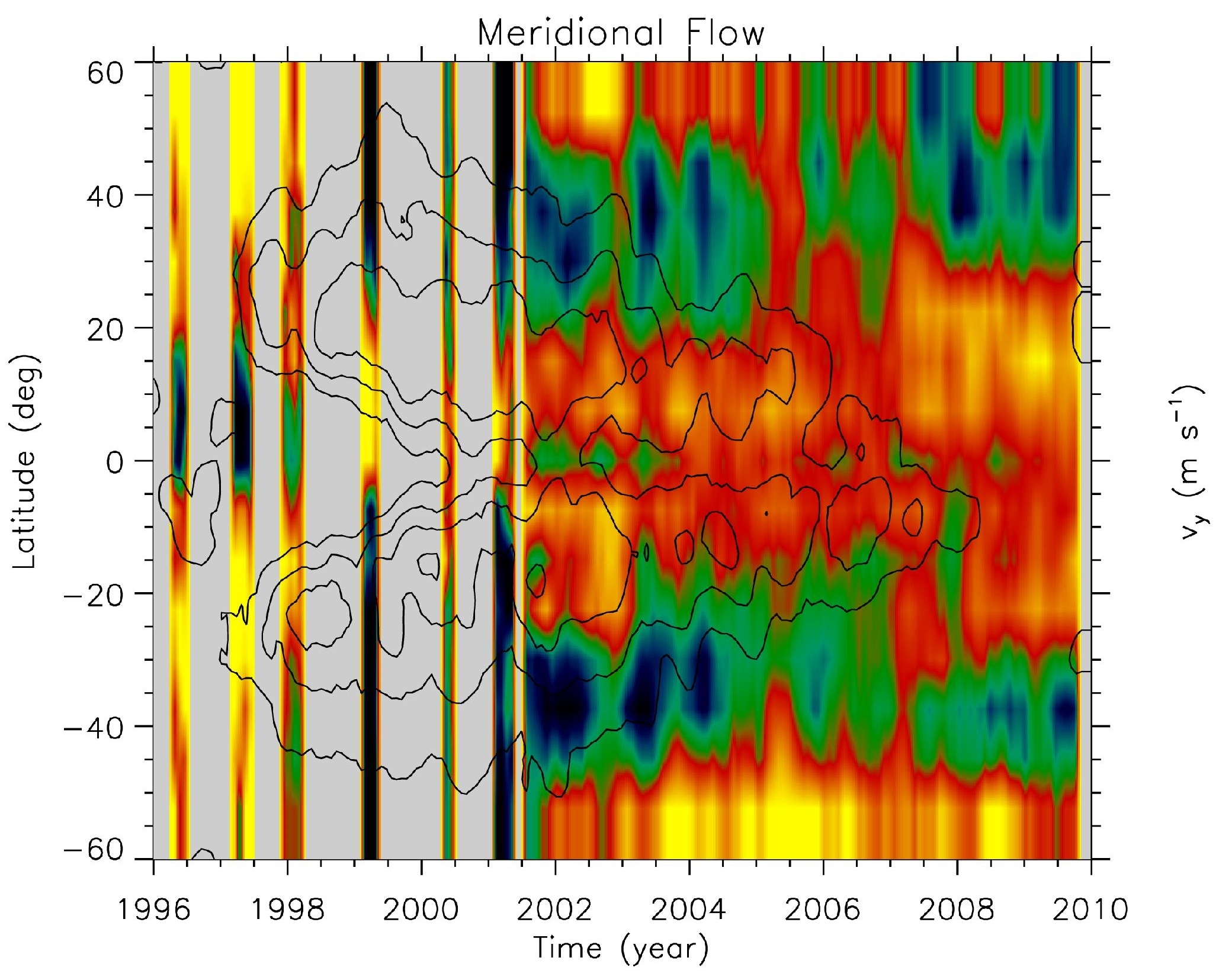}
\caption{Komm et al. (2011) recovered zonal (left) and meridional (right) flows as a function of time and latitude for a depth of 10.2- 15.8 Mm. These results were determined from SOHO MDI (before mid-2001) and GONG (after mid-2001) data. Overlaid contours show average magnetic field strength. }
\label{fig:Komm2011} 
\end{figure}

The extent of the hemispheric coupling as determined by helioseismology is as follows.  Zonal flows are seen as bands of faster and slower E-W flows that appear years prior to the appearance of activity on the solar surface.  It is often thought that the flow patterns are caused by enhanced cooling by magnetic fields.  Meridional flows are considered in many dynamo models as the crucial ingredient which sets the rate at which the toroidal magnetic band (and sunspots) move equatorward. \citet{2011JPhCS.271a2077K} determined the zonal and meridional flows as a function of latitude for cycle 23 (years 1996-2010), see Fig.~\ref{fig:Komm2011}. In the right panel of Fig.~\ref{fig:Komm2011}, note how the meridional flow at 10-15 Mm in the Northern hemisphere weakens in 2005 at 35 N latitude (seen as a break in green color) just before the Northern surface magnetic contour disappears in 2006.  Similarly, the Southern hemisphere shows this behavior 2 years later in 2007 at 35 S latitude (again, a break in the green color) just before the Southern magnetic contour disappears in 2008.  This $\sim2$ year hemispheric phase lag observed in both the surface magnetism and the meridional flow is tantalizing.

\begin{figure}
\includegraphics[width=1.00\textwidth]{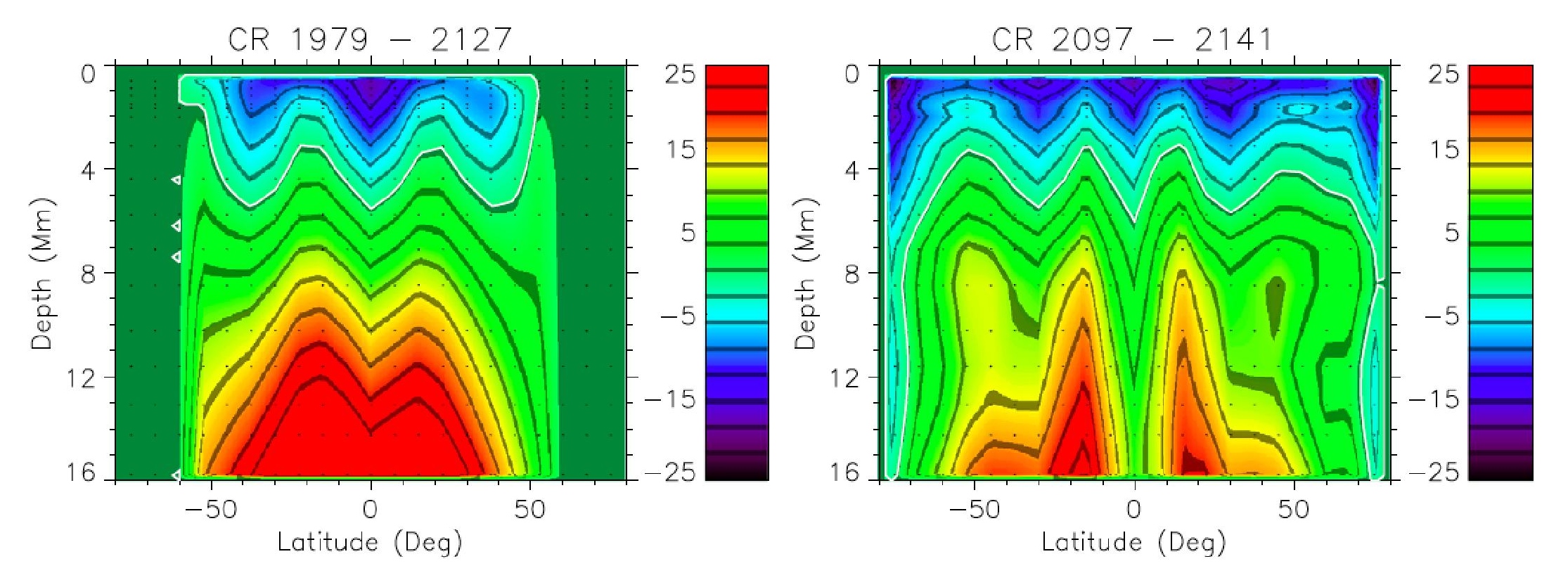}
\caption{Komm et al. (2014) recovered zonal flows as a function of latitude and depth.  The zonal flow from GONG (left) is averaged over 149 Carrington rotations and from HMI (right) is averaged over 35 Carrington rotations.  The white contour indicates a zero velocity.  Errors are less than 0.1 m s$^{-1}$ for GONG and less than 1.0 m s$^{-1}$ for HMI.  The poleward branch of the torsional oscillations is stronger in the S by ~6 m s$^{-1}$ at a depth of 10-13 Mm during Cycle 23.}
\label{fig:Komm2014} 
\end{figure}

More recently, \citet{2014SoPh..tmp...29K} investigated the behavior of the zonal flows as a function of latitude for the time period of 2001 - 2013 from the surface to a depth of 16 Mm using GONG and HMI.  Many hemispheric differences are evident in the zonal flows.  For example, see Fig.~\ref{fig:Komm2014} showing the poleward branch of the zonal flow (at 50 degrees) is 6 m s$^{-1}$ faster in the S than the N at a depth of 10 $-$ 13 Mm during cycle 23. In addition, \citet{2013ApJ...774L..29Z} detected multiple cells in each hemisphere in the meridional circulation using acoustic travel-time differences. The double-celled profile shows a significant hemispheric asymmetry (see panels (d) and (e) in Fig.~\ref{fig:Zhao2013}) in a range of latitudes. The profile asymmetry could be due to a phase lag in the hemispheres: does the meridional profile as a function of depth in the Southern hemisphere in 2013 look like the profile did in the Northern hemisphere two years earlier?

It is possible that a perturbation of meridional flow (presumably by convection since the meridional flow is a weak flow strongly driven by convection) in one hemisphere (but not in the other) sets a phase lag between the migration of activity belts, and hence, the sunspot production, that persists for years.   Actively searching for correlated hemispheric asymmetric signatures in flows at depth and magnetic field distributions on the surface may provide insight as to which ingredients of the dynamo set the length and amplitude of the sunspot cycle.  For an in-depth discussion of N-S hemispheric asymmetry from an observational perspective as compared to the results from dynamo simulations, see Norton et al. (2014) in this volume.

\section{Mean field modelling of solar torsional oscillations}

Several authors \citep{Durney:2000, Covas:etal:2000, Bushby:2006, Rempel:2006:dynamo}
have developed theoretical models of torsional oscillations, the observational aspects of which are described in Section \ref{sec:2.3}, by assuming
that the torsional oscillations are driven by the Lorentz force of the Sun's cyclically
varying magnetic field, which is associated with sunspot
cycle. If this is true, then one would expect
the torsional oscillations to follow the sunspot cycles.
The puzzling fact, however, is that the torsional oscillations of
a cycle begin a couple of years before the sunspots of that
cycle appear and at a latitude higher than where
the first sunspots are subsequently seen.  At first sight, this
looks like a violation of causality. How does the effect precede the cause?
The following section provides a possible explanation using the Nandy-Choudhuri hypothesis \citep{Nandy:Choudhuri:2002}
in the framework of flux transport dynamos.
\subsection{Features of solar torsional oscillations and their possible explanations}
\label{sec:obstorosc}
We now summarize some of
the other important characteristics of torsional oscillations,
which a theoretical model should try to explain.

\begin{enumerate}
  \item Apart from the weaker equatorward-propagating branch which moves with the sunspot
belt after the sunspots start appearing, there is also a
stronger poleward-propagating branch at high latitudes. The poleward branch lasts for about 9 years whereas the equatorward branch lasts for about 18 years (e.g. see Fig. \ref{tors-lat}).
  \item The amplitude of the torsional oscillations near the surface is of the order of 4 nHz (20 ms$^{-1}$) when averaged over all latitudes (e.g. see Fig. \ref{tors-lat}).
  \item The torsional oscillations seem to be present throughout the convection zone, though they appear more intermittent and less coherent as we go deeper down into the convection zone \citep[see Figs.\ 4, 5 and 6 in ][and Fig. \ref{tors-depth} here]{2005ApJ...634.1405H}.
  \item In the equatorward-propagating branch at low latitudes, the torsional oscillations at the surface seem to have a phase lag of about 2\,yr compared to the oscillations at the bottom of the convection zone \citep[see Fig.\ 7 in ][and Fig. \ref{tors-depth} here]{2005ApJ...634.1405H}. The rate of upward movement appears to be about 0.05 \Rs yr$^{-1}$ or $1\,\rm m\,s^{-1}$. However, we recall from Section \ref{sec:2.3} that evidence for upward propagation is weaker in cycle 24 than observed previously.
  \item Torsional oscillation contours of constant phase are inclined at $25^\circ$ to the rotation axis. This is similar to the inclination of contours of constant rotation \citep[see Fig. \ref{HMIrot} and][]{Howe:etal:2004}.
\end{enumerate}

Let us now try to understand properties (3) and (4) listed above in some more detail.
Property (4) of the torsional oscillations seems to
suggest that the bottom of the convection zone is the source of the
oscillations, which then propagate upwards. Property (3) then
seems puzzling and contrary to common sense. One would expect
the oscillations to be more coherent near the source, becoming
more diffuse as they move upward further away from the source.
However, we know that the magnetic field is highly intermittent
within the convection zone and we need to take account of this
fact when calculating the Lorentz force due to the magnetic field.
Since the convection cells deeper down are expected to have larger
sizes, \citet{Choudhuri:2003} suggested that the magnetic field within
the convection zone would look as shown in Fig.~1 of that paper.
In Fig.~1 of \citeauthor{Choudhuri:2003} the convection granule size is a function of the pressure scale height in the solar
convection zone, the intergranular lanes are more intermittent at the
bottom of the convection zone and so is the
magnetic field which is concentrated into tubes in these lanes.
Since the velocity
perturbations associated with the torsional oscillations are
likely to be concentrated around the magnetic flux tubes, we
expect the torsional oscillations to be spatially intermittent
at the bottom of the convection zone, as seen in the observational
data (property 3). Similarly, since the magnetic field near the surface is less intermittent,
a torsional oscillation driven by the Lorentz stress would also appear
more coherent there.  This may explain the puzzling situation where
the torsional oscillations seem to become more coherent as they
move further away from the source which is at the footpoints of flux tubes
at the bottom of the convection zone.
This scenario therefore provides a natural explanation for
property (3) of the torsional oscillations, as listed above. Accordingly, let us
assume that the torsional oscillation gets initiated in
the lower footpoints of the vertical flux tubes, where the Lorentz
force builds up due to the production of the toroidal
magnetic field. This perturbation
then propagates upward along the vertical flux tubes
at the Alfv\'en speed. If the magnetic field inside the flux
tubes within the solar convection zone (not below it) is estimated as 500 G (see Section 3 of \citet{Choudhuri:2003}),
then the Alfv\'en speed at the bottom of the
convection zone is of the order of 300 cm s$^{-1}$ and the
Alfv\'en travel time from the bottom
to the top turns out to be exactly of
the same order as the phase delay of torsional oscillations
at the surface compared to the oscillations at the bottom of
the convection zone. This may be an explanation for the property (4). In the next subsection, we shall try to incorporate these ideas into a mean field dynamo model of the solar cycle.

\subsection{Numerical Modelling efforts}

We focus on the scenario where the Lorentz and MAxwell stresses on the flows in the solar interior give rise to the torsional oscillations. Poloidal flows, like the meridional circulation, are also affected. This is important because the meridional circulation is believed to set the clock for the 11yr sunspot cycle. The magnetic feedback on the flows is one of the important non linearities of the solar dynamo. Some authors have proposed that such feedback on differential rotation alone could cause a modulation of solar cycle strengths and lead to grand minima like episodes \citep{Kitchatinov:etal:1999}.

Mean field models of the solar dynamo have been in existence for over two decades now and come in two main flavors: the $\alpha \Omega$, interface, and the flux transport dynamo. Depending on the values of parameters like turbulent diffusivity as compared to meridional circulation speed, the dynamo can also be characterized into advectively or diffusively dominated dynamo regimes \citep{Jiang:etal:2007, Yeates:etal:2008}. Such a mean field model of the solar dynamo can then be combined with the Reynolds-averaged Navier-Stokes (NS) equation, after including the effects of the Lorentz force and Maxwells stresses, to solve for the torsional oscillations. Several authors \citep{Covas:etal:2000, Covas:etal:2004, Chakraborty:etal:2009} have solved for only the azimuthal component of the mean velocity, $v_{\phi}$ along with the dynamo equations while others \citep{Rempel:2007} have solved for the evolution of entropy in addition to all three components
of the mean velocity.  Torsional oscillation models also differ in relation to the magnetic feedback term appearing in the NS equation. In some cases \citep{Rempel:2007, Chakraborty:etal:2009, Covas:etal:2004} only a Lorentz feedback due to the mean magnetic field (macro-feedback), has been used whereas in other cases \citep{Kitchatinov:etal:1999, Kueker:etal:1999} only the Maxwell's stress due to the fluctuating magnetic field (micro-feedback) has been modeled. This is achieved using $\Lambda$-effect formulation of the non-diffusive component of the Reynolds stresses \citep{Kitchatinov:Ruediger:1993} and quenching the $\Lambda$ coefficients using an algebraic dependence on the mean magnetic field. The saturation of the mean magnetic field in the above case has been achieved using two different mechanisms: In the first mechanism the $\Lambda$ coefficients are quenched along with either the kinetic helicity or the $\alpha$ effect (achieved through the presence of super equipartition fields). The second mechanism was employed by \citet{Covas:etal:2004} who demonstrated that saturation of the solar dynamo can be obtained by means of the Lorentz
feedback in the $\phi$ component of the NS equation without requiring any explicit $\alpha$-quenching.

One of the popular models used for an explanation of the sunspot cycle is the flux transport
dynamo, in which the meridional circulation carries the toroidal
field produced from differential rotation in the tachocline
equatorward and carries the poloidal field produced
by the Babcock--Leighton mechanism at the surface poleward \citep{Wang:Sheeley:1991, Choudhuri:etal:1995}.
Since the differential rotation is stronger at higher latitudes
in the tachocline than at lower latitudes,
the inclusion of solar-like rotation tends to produce a strong
toroidal field at high latitudes rather than at the latitudes where
sunspots are seen \citep{Dikpati:Charbonneau:1999, Kueker:etal:2001}. \citet{Nandy:Choudhuri:2002} proposed a hypothesis to
overcome this difficulty. According to their hypothesis,
the meridional circulation penetrates slightly below the bottom
of the convection zone and the strong toroidal field produced at the
high-latitude tachocline is pushed by this meridional circulation
into stable layers below the convection zone where magnetic buoyancy
is suppressed and sunspots are not formed.
Only when the toroidal field is brought into the
convection zone by the meridional circulation rising at lower latitudes,
does magnetic buoyancy take over and sunspots finally form. The torsional oscillation signals however cannot be buried
below the convection zone by the meridional circulation since
they can be transmitted out to the surface by Alfv\'en waves along vertical magnetic field lines, which intermittently thread the convection zone. It may be noted
that there is a controversy at the present time as to whether the
meridional circulation can penetrate below the convection
zone since arguments having been advanced both against
\citep{Gilman:Miesch:2004} and
for it \citep{Garaud:Brummel:2008}. 
The detailed dynamo model of \citet{Chatterjee:etal:2004} was based on this Nandy-Choudhuri hypothesis, which provided the correct sunspot emergence latitudes and phase relation between the low-latitute toroidal field and the polar fields at the surface. 

\begin{figure}
\includegraphics[width=0.5\textwidth, clip, trim=1cm 6cm 1cm 9cm]{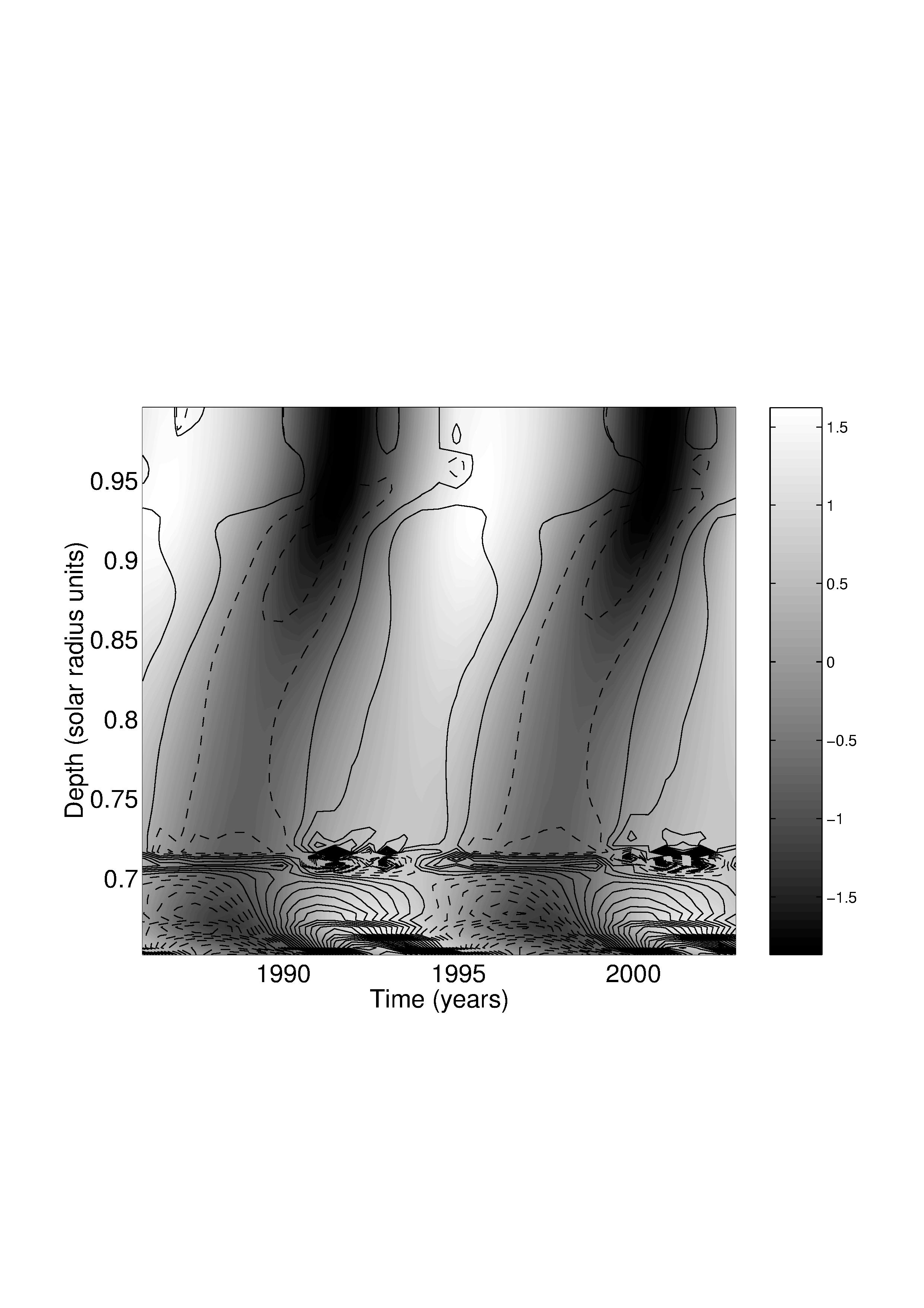}
\includegraphics[width=0.5\textwidth, clip, trim=1cm 6cm 1cm 9cm]{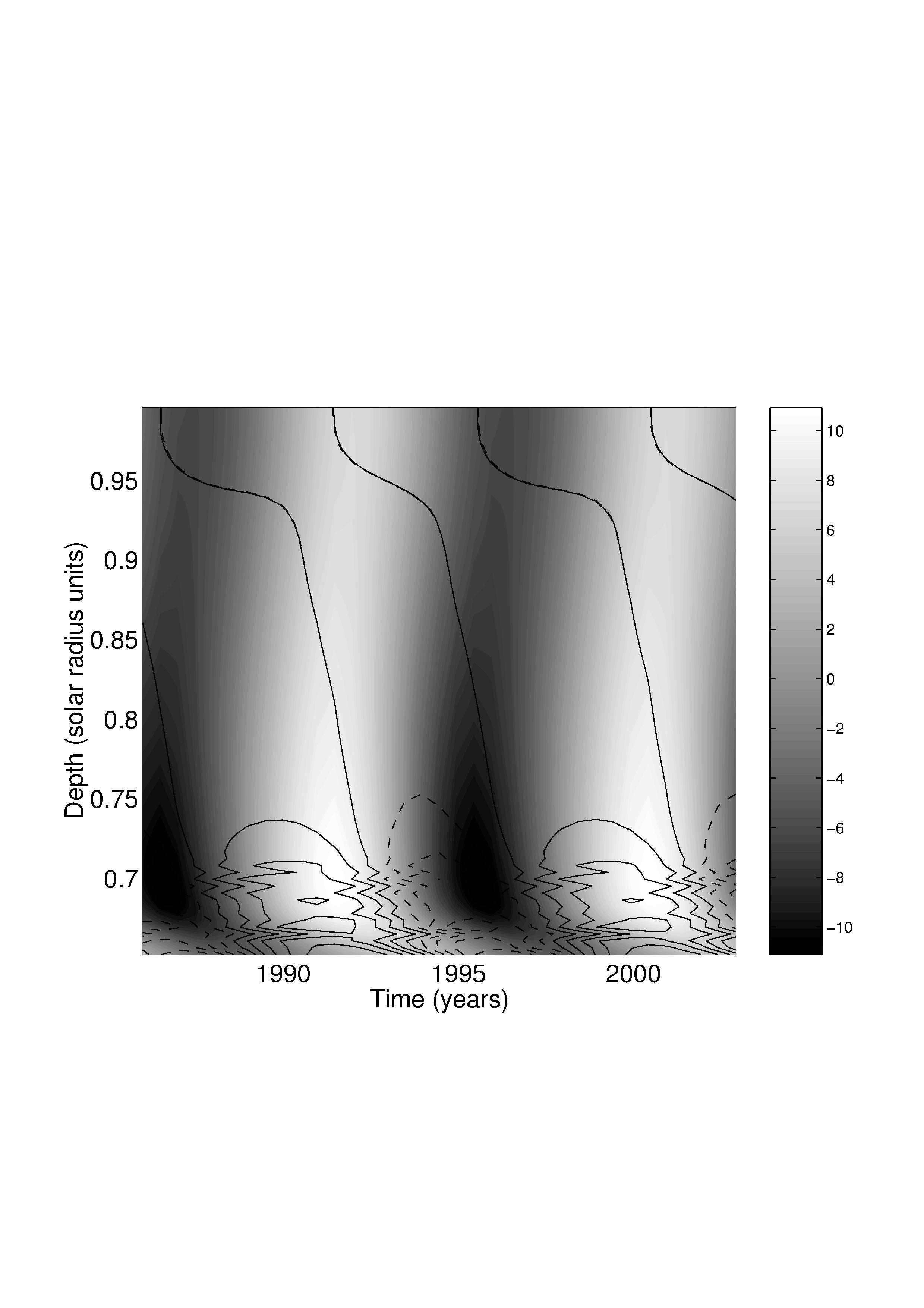}
\caption{Theoretical torsional oscillations ($\vp$ in m s$^{-1}$) as functions of depth and time at
latitudes $20^{\circ}$ (left) and $70^{\circ}$ (right).  The plot for latitude $20^{\circ}$ compares
very well with Fig.~4(D) of \citep{2002Sci...296..101V}, Fig.~7 \citep{2005ApJ...634.1405H}, or Fig. \ref{tors-depth} here.
The contours indicate the Lorentz force $({\bf F}_L)_{\phi}$, the solid and
dashed lines indicating positive and negative values.}\label{fig:chakfg2}
\end{figure}

Another conjecture proposed by \citet{Chakraborty:etal:2009} was that the torsional oscillations are initiated in the lower layers of the solar convection zone where toroidal flux tubes are formed due to differential rotation and are propagated upwards by Alfv\'en waves. The torsional oscillation signal therefore reaches the solar surface much ahead of the sunspot-forming toroidal magnetic field which is still buried by the downward meridional flow in the stable layers.  These authors tried to model torsional oscillations using the \citet{Chatterjee:etal:2004} 
dynamo model after coupling it to the equation for the mean rotational velocity incorporating a very simple but insightful averaging of the Lorentz force feedback term. 
The $\phi$ component of the Navier--Stokes equation,
is
\begin{equation}\label{eq[ns]}
    \rho \left\{ \frac{\pa \vp}{\pa t} + D_u [\vp] \right\} =
D_{\nu} [\vp] + ({\bf F}_L)_{\phi},
\end{equation}
where
\begin{equation}\label{eq[advection]}
    D_u [\vp] = \frac{v_r+v_{\rm alf}}{r} \frac{\pa}{\pa r} (r \vp)
+ \frac{v_{\theta}}{r \sin \theta} \frac{\pa}{\pa \theta} (\sin \theta \vp)
\end{equation}
is the term corresponding to advection by the meridional circulation $(v_r, v_{\theta})$, and $v_{\rm alf}$ is a constant upward velocity of 
$300$ cm s$^{-1}$, estimated in Section \ref{sec:obstorosc},
to account for the upward transport by Alfv\'en waves when solving our basic equation Eq.~\ref{eq[ns]}. Note that this additional $v_{\rm alf}$ 
does not represent any
actual mass motion and does not have to satisfy the continuity equation, unlike the meridional circulation.
\begin{equation}\label{eq[diffusion]}
    D_{\nu} [\vp] = \frac{1}{r^3} \frac{\pa}{\pa r} \left[ \nu \rho r^4 \frac{\pa}{\pa r} \left( \frac
{\vp}{r} \right) \right] + \frac{1}{r^2 \sin^2 \theta} \frac{\pa}{\pa \theta} \left[ \nu \rho
\sin^3 \theta \frac{\pa}{\pa \theta} \left( \frac{\vp}{\sin \theta} \right) \right]
\end{equation}
is the diffusion term, and $({\bf F}_L)_{\phi}$ is the $\phi$ component of the Lorentz force.
The kinematic viscosity $\nu$ is primarily due to turbulence within the convection
zone and is expected to be equal to the magnetic diffusivity.
If the magnetic field in our model is assumed
to have the standard form
\begin{equation}\label{eq[mag field]}
    {\bf B} = B (r, \theta, t){\bf e}_{\phi} + \nabla \times [A(r, \theta, t){\bf e}_{\phi}],
\end{equation}
then the Lorentz force is given by the Jacobian
\begin{equation}\label{eq[lorentz]}
    4 \pi ({\bf F}_L)_{\phi} = \frac{1}{s^3} J \left( \frac{s B_{\phi},
s A }{r, \theta} \right),
\end{equation}
where $s= r \sin \theta$.
We, however, have to take some special care in averaging the term in 
Eq.~\ref{eq[lorentz]}, since this is the only term in our equations which is quadratic in the 
basic variables $(A, B, \vp)$ and
has to be averaged differently from all the other terms due to intermittency of the magnetic field in the convection zone. The $\phi$ component of the Lorentz
force primarily comes from the radial derivative of the magnetic stress $B_r \Bp /4\pi$.
This stress arises when $B_r$ is stretched by differential rotation to produce $\Bp$
and should be non-zero only inside the flux tubes.
We assume that $B_r, \Bp$ are the mean field values, whereas $(B_r)_{\rm ft},  (\Bp)_{\rm ft}$
are the values of these quantities inside flux tubes.  If $f$ is the filling factor, which is essentially the fractional volume occupied by flux tubes,
then we have $B_r = f (B_r)_{\rm ft}$
and $\Bp = f (\Bp)_{\rm ft}$, on assuming the same filling factor for both components
for the sake of simplicity. It is easy to see that the mean Lorentz stress would be
\begin{equation}\label{eq[lorentz stress]}
    f \frac{(B_r)_{\rm ft}(\Bp)_{\rm ft}}{4 \pi} = \frac{B_r \Bp}{4 \pi f}.
\end{equation}
This suggests that the correct mean field expression for $({\bf F}_L)_{\phi}$ should
be given by the Eq.\ref{eq[lorentz]} divided by $f$.   As pointed out by \citet{Chatterjee:etal:2004}, the only non-linearity in our equations comes from the critical magnetic field $B_c$ above which the toroidal field at the bottom of the convection zone is supposed to be unstable due to magnetic buoyancy.  \citet{Jiang:etal:2007}
found that we have to take $B_c = 108$ G (which is the critical value of the mean toroidal
field and not the toroidal field inside flux tubes) to ensure that the poloidal field at
the surface has correct values.  Once the amplitude of the magnetic field gets fixed in this manner, we find that the amplitude of
the torsional oscillations matches observational values only for a particular value of the filling factor $f$.  The theoretical model of torsional
oscillations proposed by \citet{Chakraborty:etal:2009} thus
allows us to infer the filling factor of the magnetic field in the lower layers of the
convection zone. Even though the conjecture of \citeauthor{Chakraborty:etal:2009} looks very elegant, upon actual calculation they obtained a large phase lag of $\sim6$ years instead of the observed 2 years between the onset of the low-latitude branch of the torsional oscillation and the first appearance of sunspots of the cycle. However, the radial dependence of the torsional signal was very close to the observed patterns and is shown in Fig. \ref{fig:chakfg2}. It is clear in the
plot of the torsional oscillation at a latitude of $20^{\circ}$ (left panel of Fig. \ref{fig:chakfg2}) that the Lorentz force is concentrated in the tachocline at 0.7\Rs,
where the low-latitude torsional oscillations are launched to propagate upward. The plot for latitude
$20^{\circ}$ shows that the amplitude of the torsional oscillations becomes
larger near the surface due to the perturbations propagating into regions of
lower density, which is consistent with observational
data. The physics of the high-latitude branch (right panel of Fig. \ref{fig:chakfg2}) is, however, very different, with the Lorentz
force contours indicating a downward propagation and not a particularly strong concentration
at the tachocline.  As the poloidal field sinks with the downward meridional circulation
at the high latitudes, the latitudinal shear $d \Omega/ d \theta$ in the convection
zone acts on it to create the toroidal component and thereby the Lorentz stress.
With the downward advection of the poloidal field, the region of Lorentz stress
moves downward.

 Yet another baffling observation which has eluded modelers is the systematic deviation of the isolines of constant phase of the torsional oscillation from the Taylor-Proudman state, where the fluid velocity should be uniform along contours parallel to the rotation axis, as evident from Fig.\,3B of \cite{2002Sci...296..101V}. \cite{Howe:etal:2004} first pointed out that lines of constant phase are inclined at $25^\circ$ to the rotation axis, similar to the isorotation contours. \cite{Spruit:2003} suggested a thermal origin for the low-latitude branch of the torsional oscillation due to enhanced radiative losses in the active region belts. \cite{Rempel:2007} showed that only a localized thermal forcing can lead to such deviation from the Taylor-Proudman state. However since the thermal forcing is only concentrated near the surface, it does not explain the observational fact that torsional oscillations encompass almost the entire convection zone. Earlier, \citet[][see Fig.~10 of that paper]{Rempel:2006:dynamo} showed that applying a surface cooling function confined to sunspot emerging latitudes can give rise to the low-latitude branch of the torsional oscillation where as the magnetic forcing is responsible for the high-latitude branch. According to our knowledge none of the authors have yet obtained the correct phase relation between the low-latitude band of the torsional oscillation and the sunspot migration. The precedence of the low-latitude branch over the first appearance of sunspots is still a mystery. 

Note that, all the discussion above are for mean field models which give rise to regular solar cycle amplitudes for a relatively weak magnetic feedback. An exception is the model of \cite{Kitchatinov:etal:1999} which has a strong magnetic feedback and gives rise to significant modulation in the strengths of successive solar cycles. In reality, the solar cycle amplitudes are irregular or stochastic in nature. Naturally this would give rise to an irregular magnetic feedback on the solar flows and the resulting observed torsional oscillations. The irregular magnetic feedback can not only affect the amplitude of torsional oscillations but also the mean rotation rate of the Sun. It may be important to define the mean rotation rate as a solar cycle averaged quantity rather than a very long temporal average 
\citep[see][]{2013ApJ...767L..20H}. Thus, modelling the solar torsional oscillations still remains a very challenging problem. 

\section{Was cycle 23 unusual?}

\begin{figure}
  \centering
  \includegraphics[width=0.6\textwidth, clip=true]{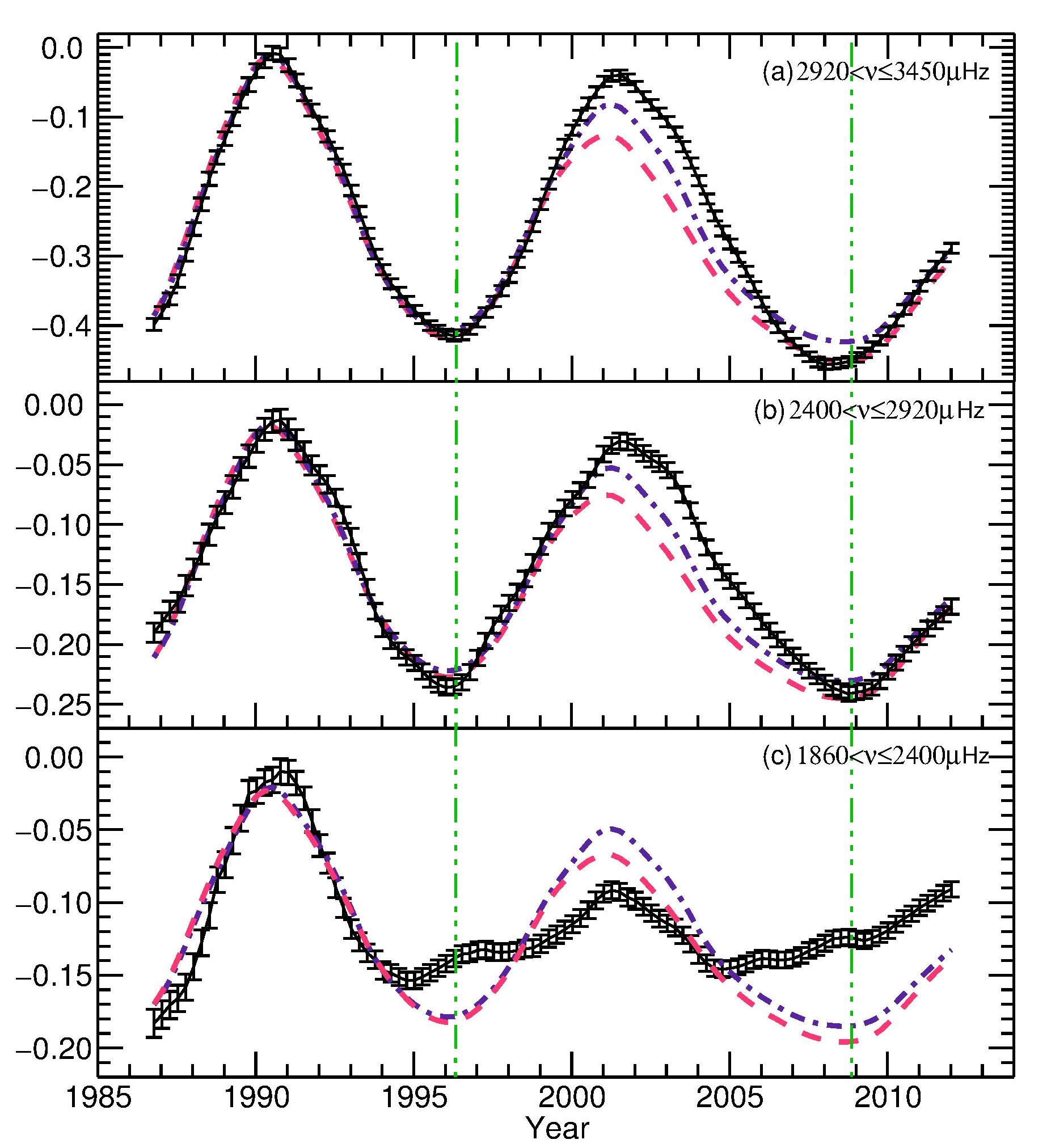}\\
  \caption{Smoothed frequency shifts as a function of time observed in three frequency ranges (see captions). The frequency of the oscillations were obtained from 365\,d Sun-as-a-star BiSON data. An average was taken over modes with $0\le l \le 2$. This figure is adapted from \citet{2012ApJ...758...43B}. Scaled, shifted and smoothed versions of the 10.7cm flux (blue dot-dashed line) and the ISN (red dashed line) are plotted for comparison purposes. }\label{figure[smoothed_shifts]}
\end{figure}

The minimum that preceded cycle 24 was considered to be unusually long and quiet, even for a solar minimum. What can the solar oscillations tell us about the structure of the Sun's magnetic field during this time? \citet{2012ApJ...758...43B} examined the frequency dependence of the frequency shifts observed in Sun-as-a-star data during the last two solar cycles. Fig.\ref{figure[smoothed_shifts]} shows the frequency shifts of low-$l$ ($\le 2$) modes in three frequency ranges. The shifts have been smoothed to remove the quasi-biennial variation \citep[e.g.][]{2010ApJ...718L..19F}. Fig. \ref{figure[smoothed_shifts]} shows that the low-frequency modes behave unexpectedly, not just during the recent unusual solar minimum but for the entirety of cycle 23, experiencing little to no frequency shift during this time. Although it is expected that the low-frequency modes experience a smaller shift in frequency than the high-frequency modes (see Section \ref{section[freq dependence of shifts]}), a comparison with the shifts observed in cycle 22 highlights the discrepancy. More precisely, while the behaviour of the high and intermediate ranges is consistent between cycles 22 and 23, the behaviour of the low-frequency modes changes. This can be explained in terms of the upper turning points of the modes, which as we have said previously, are dependent on the frequency of the mode. If we consider the perturbation responsible for the solar-cycle frequency shifts to be a near-surface magnetic layer. In cycle 22 the upper turning points of the low-frequency modes must lie within the magnetic layer, but in cycle 23 the upper turning points of the low frequency modes must lie beneath the layer, meaning their frequencies are not perturbed by it. These results, therefore, imply a thinning of the magnetic layer (or a change in the upper turning points with respect to the layer). \citet{2012ApJ...758...43B} therefore infer that the magnetic layer must be positioned above $0.9965R_\odot$ in cycle 23.

\begin{figure}[!ht]
\centering
\includegraphics*[width=8.0cm]{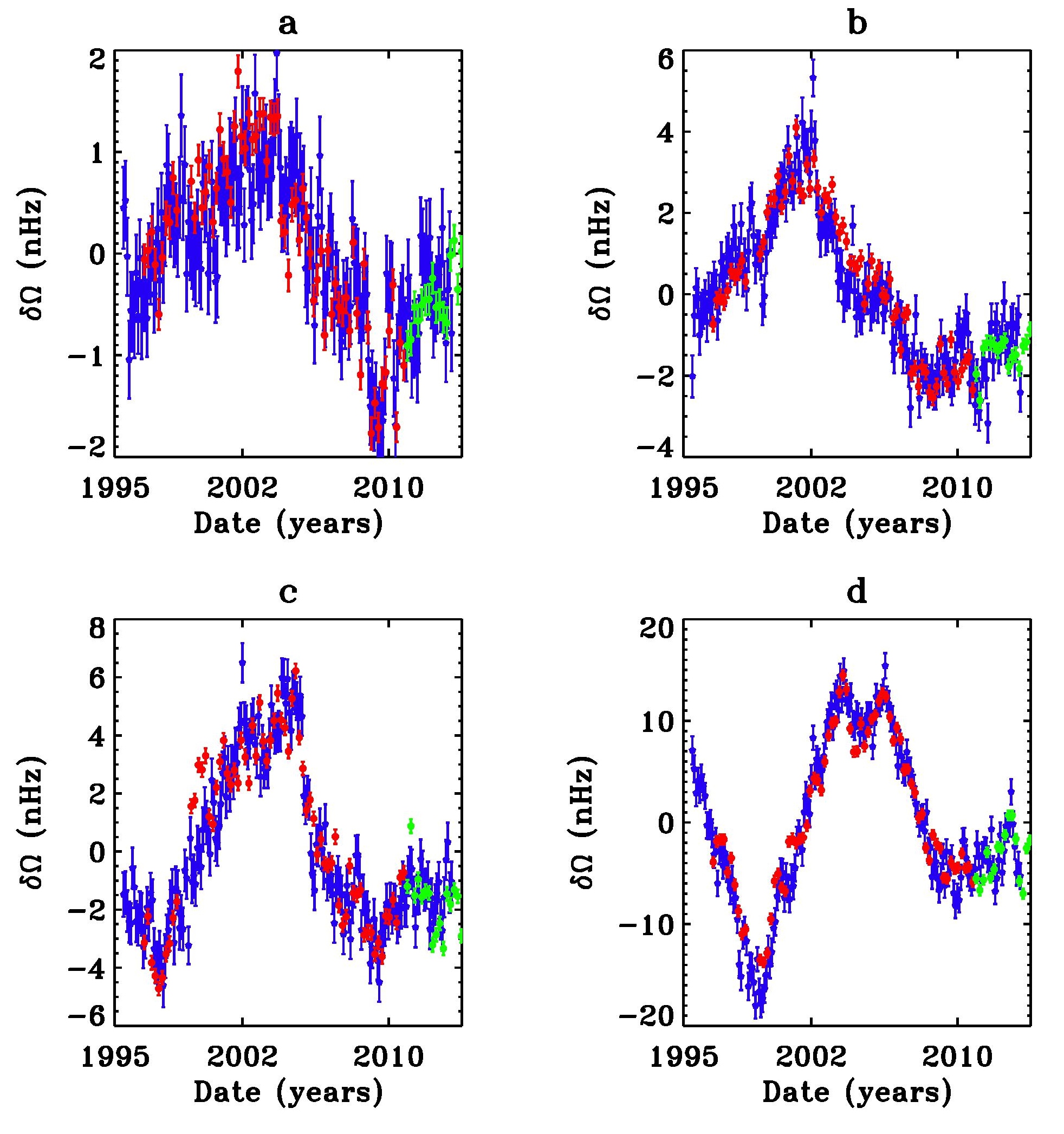}
\caption{Rotation rate from Regularized Least Squares inversions (after subtraction of a temporal mean) at latitudes (from left to right) $45^\circ$,
$56^\circ$, $68^\circ$ and $79^\circ$, at radius $0.99R$, inferred using
an RLS inversion. Colours indicate the
data used: GONG (blue), MDI (red) and HMI (green).}\label{hilat}
\end{figure}

A change in behaviour has also been observed in the torsional oscillation. In Fig.~\ref{tors-lat} it appears that the high-latitude poleward-propagating
spin-up, which was prominent in the 2000-2006 epoch, is absent in the
present cycle. It has been speculated that the high-latitude branch is a
precursor of the following solar cycle, and that its absence at the present
time may indicate that cycle 25 may be delayed, weak, or non-existent.
Another way to look at the data is to plot the inferred rotation at a single
location, as a function of time (Fig.~\ref{hilat}). It is clear from this
representation that the rotation rate at mid- and high-latitudes has been
increasing in the past 2-3 years, but more weakly than in the previous cycle.
At mid-latitudes, it is also evident that the rotation rate dropped to a lower
level than in the previous cycle, so even the weak increase is starting from a
lower base \citep{2013ApJ...767L..20H}.

\section{Summarizing remarks}

The Sun's magnetic field, which is generated in the solar interior, varies on a time scale of 11\,yrs from minimum to maximum and back again. Measures of the Sun's surface and atmospheric magnetic field are abundant. In order to really understand the interior of the Sun we use the Sun's natural oscillations. The frequencies, powers and lifetimes of these oscillations are dependent on the strength of the Sun's magnetic field, with the most significant influence on the oscillations arising from a near-surface perturbation. In fact, it has been shown that the change in frequency of the oscillations is tightly correlated with the surface magnetic field, once latitudinal distribution is taken into account. Evidence for a magnetic field deeper within the solar interior was hard to come by. However, there is now tentative evidence for solar cycle variations in the second-ionization zone of helium and at the base of the convection zone. Solar cycle variations in the dynamics of the solar interior have been far more forthcoming. For example, the link between flows and active regions, such as sunspots, have been well studied and reveal strong outflows around sunspots. The torsional oscillation has been shown to exist over a substantial fraction of the convection zone. The observational constraints of the torsional oscillations have proven to be key in recent developments in mean field dynamo models. This is particularly pertinent since one of the main aims in any solar cycle study is to improve our understanding of the mechanism by which the Sun's magnetic field is generated and maintained. Any model of the solar dynamo must explain observational features, including the North-South asymmetry, which is apparent, not only in surface measures of the Sun's magnetic field but in the flows seen in the solar interior. Finally it has been shown that, from a helioseismic standpoint the Sun's activity appears to be changing from one cycle to the next. We must remember here that we only have two 11-yr solar cycles for comparison. Although the BiSON data do go back to cycle 21, the fill of the data is poor as the full 6-site network was not established until the early 1990s, meaning that it is hard to make helioseismic inferences about this cycle. It is, therefore, difficult to say, purely in helioseismic terms, which cycle is behaving unusually, or indeed if the behaviour is unusual at all: 11\,yr only covers half a 22-yr Hale cycle. One can, therefore, only conclude that the helioseismic data imply that the behaviour of the Sun has changed. Hopefully these and other issues surrounding the Sun's magnetic field will become clearer as we continue to observe the Sun through cycle 24 and beyond.

\begin{acknowledgements} The paper was stimulated by the workshop ``The solar activity cycle: physical causes and consequences". It is a pleasure to thank Andr\'e Balogh, Hugh Hudson, Kristof Petrovay, Rudolf von Steiger  and the  International Space Science Institute for financial support, excellent organization and hospitality. This work utilizes Birmingham Solar Oscillations Network data which is run by School of Physics and Astronomy, University of Birmingham. This work utilizes GONG data obtained by the NSO Integrated Synoptic Program (NISP), managed by the National Solar Observatory, which is operated by AURA, Inc. under a cooperative agreement with the National Science Foundation. The data were acquired by instruments operated by the Big Bear Solar Observatory, High Altitude Observatory, Learmonth Solar Observatory, Udaipur Solar Observatory, Instituto de Astrof\'{\i}sica de Canarias, and Cerro Tololo Interamerican Observatory. {\it SOHO} is a mission of international cooperation between ESA and NASA. The Solar Oscillations Investigation (SOI) involving MDI was supported by NASA grant NNX09AI90G to Stanford University. The National Center for Atmospheric Research is sponsored by the National Science Foundation. HMI data courtesy of NASA and the HMI consortium; HMI is supported by NASA contract NAS5-02139 to Stanford University. NSO/Kitt Peak data used here were produced cooperatively by NSF/NOAO, NASA/GSFC, and NOAA/SEL; SOLIS data are produced cooperatively by NSF/NSO and NASA/LWS. A-MB thanks the Institute of Advanced Study, University of Warwick for their support. RH acknowledges computing support from the National Solar Observatory. A. Norton is supported by NASA Contract NAS5-02139 (HMI) to Stanford University. We thank N. Featherstone and R. Komm for their help. 
\end{acknowledgements}

\bibliographystyle{aps-nameyear}
\bibliography{ISSI_helio_chapter}

\end{document}